\documentclass{article}
\usepackage{arxiv}
\usepackage[utf8]{inputenc} % allow utf-8 input
\usepackage[T1]{fontenc}    % use 8-bit T1 fonts
\usepackage{url}            % simple URL typesetting
\usepackage{booktabs}       % professional-quality tables
\usepackage{amsfonts}       % blackboard math symbols
\usepackage{nicefrac}       % compact symbols for 1/2, etc.
\usepackage{microtype}      % microtypography
\usepackage{graphicx}
\usepackage{epstopdf, epsfig}
\usepackage{subfigure}
\usepackage{amsmath,amssymb}
\usepackage{lineno}
\usepackage{algorithm2e}
\usepackage{tabularx}
\usepackage{easyReview}%added by Tak.

\title{OpenMetBuoy-v2021: an easy-to-build, affordable, customizable, open source instrument for oceanographic measurements of drift and waves in sea ice and the open ocean}
\author{
   Jean Rabault\\
   Norwegian Meteorological Institute\\
   \textit{jean.rblt@gmail.com}\\
   \And
   Takehiko Nose\\
   The University of Tokyo\\
   \textit{tak.nose@edu.k.u-tokyo.ac.jp}\\
   \And
   Gaute Hope\\
   Norwegian Meteorological Institute\\
   \textit{gaute.hope@met.no}\\
   \And
   Malte Muller\\
   Norwegian Meteorological Institute\\
   \&
   University of Oslo \\
   \textit{maltem@met.no}\\
   \And
   Oyvind Breivik\\
   Norwegian Meteorological Institute\\
   \textit{oyvindb@met.no}\\
   \And
   Joey Voermans\\
   The University of Melbourne\\
   \textit{jvoermans@unimelb.edu.au}\\
   \And
   Lars Robert Hole\\
   Norwegian Meteorological Institute\\
   \textit{lrh@met.no}\\
   \And
   Patrik Bohlinger\\
   Norwegian Meteorological Institute\\
   \textit{patrikb@met.no}\\
   \And
   Takuji Waseda\\
   The University of Tokyo \& \\
   Japan Agency for Marine-Earth Science and Technology \\
   \textit{waseda@k.u-tokyo.ac.jp }\\
   \And
   Tsubasa Kodaira\\
   The University of Tokyo\\
   \textit{kodaira@edu.k.u-tokyo.ac.jp }\\
   \And
   Tomotaka Katsuno\\
   The University of Tokyo\\
   \textit{katsuno@s.otpe.k.u-tokyo.ac.jp }\\
   \And
   Mark Johnson\\
   The University of Alaska, Fairbanks\\
   \textit{majohnson@alaska.edu }\\
   \And
   Graig Sutherland\\
   Environment Climate Change Canada, Dorval\\
   \textit{graigory.sutherland@canada.ca}\\
   \And
   Malin Johanson\\
   UiT The Arctic University of Norway\\
   \textit{malin.johansson@uit.no}\\
   \And
   Kai Haakon Christensen\\
   Norwegian Meteorological Institute\\
   \textit{kaihc@met.no}\\
   \And
   Adam Garbo \\
   Carleton University, Ottawa, Ontario, Canada \\
   \textit{adam.garbo@carleton.ca}\\
   \And
   Atle Jensen\\
   The University of Oslo\\
   \textit{atlej@math.uio.no}\\
   \And
   Olav Gundersen\\
   The University of Oslo\\
   \textit{olavgun@math.uio.no}\\
   \And
   Aleksey Marchenko\\
   The University Center in Svalbard\\
   \textit{aleksey.marchenko@unis.no}\\
   \And
   Alexander Babanin\\
   The University of Melbourne\\
   \textit{a.babanin@unimelb.edu.au}\\
}

\begin{document}
\maketitle

%\linenumbers

\begin{abstract}
  There is a wide consensus within the polar science, meteorology, and oceanography
  communities that more in-situ observations of the ocean, atmosphere, and sea ice,
  are required to further improve operational forecasting model skills.
  Traditionally, the volume of such measurements has been limited by the high cost of
  commercially available instruments. An increasingly attractive solution to this cost
  issue is to use instruments produced in-house from open source hardware, firmware, and
  post processing building blocks. In the present work, we release the next
  iteration of the open source drifter and waves monitoring instrument of Rabault et. al. (see "An open source, versatile, affordable waves in ice instrument for scientific measurements in the Polar Regions", Cold Regions Science and Technology, 2020), which follows these solution aspects. The new design is both
  significantly less expensive (typically by a factor of 5 compared to our previous,
  already cost effective, instrument), much easier to build and assemble for people
  without specific microelectronics and programming competence, more easily extendable and customizable, and two orders of magnitude more power
  efficient (to the point where solar panels are no longer needed even for long term
  deployments). Improving performance and reducing noise levels and costs compared with our previous
  generation of instruments is possible in large part thanks to progress from the electronics component industry. As
  a result, we believe that this will allow scientists in geosciences to increase by an order of
  magnitude the amount of in-situ data they can collect under a constant instrumentation
  budget. In the following, we offer 1) detailed overview of our hardware and software
  solution, 2) in-situ validation and benchmarking of our instrument, 3) full open source
  release of both hardware and software blueprints. We hope that this work, and the
  associated open source release, may be a milestone that will allow our scientific fields
  to transition towards open source, community driven instrumentation. We believe that
  this could have a considerable impact on many fields, by making in-situ instrumentation at
  least an order of magnitude less expensive and more customizable than it has been for the last
  50 years, marking the start of a new paradigm in oceanography and polar science, where
  instrumentation is an inexpensive commodity and in-situ data are easier and less expensive to
  collect.
\end{abstract}

\section{Introduction}
\label{sec:intro}

Our inability to accurately capture climatological changes of sea ice in the polar seas has created renewed interest in the dynamic interaction between sea ice and waves. This has
resulted in the last few years into a number of studies that investigate the coupling between sea ice and the ocean through theoretical considerations \citep{squire2020ocean,smith2018modelling, zhao2018three, golden2020modeling, roach2019advances, sutherland2019two, williams2017wave},
laboratory experiments
\citep{Sree2020AnES, rabault2019experiments, Li2021LaboratorySO, sutherland2017attenuation, SREE2020102233, marchenko2021laboratory},
and field experiments \citep{kohout_smith_roach_williams_montiel_williams_2020, voermans2020experimental, loken2021wave, thomson2016emerging, loken2021investigation, loken2021bringing, tc-2021-210, sutherland2016observations, rabault2017measurements, marchenko2017field, marchenko2019wave, johnson2021observing}. Despite the advances that these studies bring, there is a growing consensus that further progress in the field can  only be achieved through the collection of more observations of waves in ice.
In particular, phenomena related to floe size distribution \citep{golden2020modeling, herman2018floe, horvat2017evolution, herman2010sea}, energy dissipation due to turbulence and collisions \citep{herman2021spectral, voermans2019wave, smith2020pancake, herman2018wave}, and sea ice breakup \citep{herman2021sizes, li2021effects, ardhuin2020ice, herman2017wave},
are still unperfected modeled and understood, and advancing the state-of-the-art would require additional direct observations from the ice.

Unfortunately, one faces a number of
challenges in reproducing the complexity of wave-ice interaction either in experiments
(due to, for example, scaling issues \citep{rabault2019experiments}, and the diversity and realism of sea ice
conditions that can be obtained in the laboratory compared with the field \citep{marchenko2021laboratory}), or in
models (there, also, due to the complexity and multi-scale properties of waves in ice and
ice floe size distribution \citep{roach2018emergent, horvat2015prognostic}). Arguably, the current state of the art in wave-ice
interaction parametrization and modeling consists in a variety of (involved and
mathematically advanced) models which formulations are relatively loosely based on some
specific archetypical sea ice conditions and the associated physical mechanisms, with a
number of "tuning parameters" that are empirically fitted to experimental or laboratory
data \citep{https://doi.org/10.1002/2015JC010881, https://doi.org/10.1002/2017JC013275}. This approach is quite ad-hoc and brittle, and this is in stark contrast with
the importance of waves in ice and their impact on both weather and climate. For example,
sea-ice conditions on small spatial scales impacted by waves can have a significant impact on weather prediction skill even hundreds of kilometers away from the sea-ice edge
\citep{batrak2018atmospheric},
and sea ice conditions in the Arctic have an impact on medium range weather
forecasts over Northern Europe, and these are difficult to predict accurately at present. Maybe even more
importantly, waves in ice, sea ice, and open water areas in the Arctic and Antarctic are
coupled through a close loop feedback system: the less sea ice, the more fetch and the
higher the waves in the Arctic basin become, which leads to even more sea ice breakup and
melting \citep{https://doi.org/10.1002/2014GL059983}.

%Relatively similar challenges are also present in other sub-fields within oceanography,
%such as the study of ocean currents at sub-mesoscale. There also, collecting large amounts
%of data is necessary to allow further improvement of ocean models, which, in turn, has
%important consequences for a number of applications, including for example environmental
%protection and search and rescue operations. In the following, we will focus our
%discussion around studies of the ice drift and wave-ice interaction, but most of the
%arguments provided could be transposed to a wide range of oceanographic quantities and
%measurements.

As a consequence, having access to large datasets of high temporal and spatial resolution
direct observations of waves in ice,
is a key ingredient to improving both small scale wave-ice interaction parametrization,
large scale coupled wave-ice models, and our understanding of the weather and climate
dynamics in the polar regions. There are a number of recent works that highlight the
importance of the volume of such datasets for unveiling wave-ice interaction mechanisms.
For example, \citet{voermans2020experimental} showed that wave conditions that result in sea ice breakup events can be
quite accurately described with a simple threshold mechanism, but that in order to be able
to observe such a threshold, a relatively large aggregated dataset was necessary. Still,
this study relies on a quite small dataset compared with the sea ice extent (typically
around 7\% of the area of all oceans on Earth, averaged over one year \citep{parkinson1997earth}), and an order of
magnitude increase in the volume and resolution of the wave in ice data gathered would be
desirable to strengthen such empirical criteria and models, and to reveal more hidden
dynamics and patterns.

There are relatively few methods available for measuring waves in ice. The most common and
established method considers the deployment of instrumentation in situ, providing surface truth of the sea ice motion under the influence of waves. Such instruments typically use either an Inertial
Motion Unit (IMU, i.e. a combination of accelerometer, gyroscope, and magnetometer,
together with some data fusion processing, e.g. \citet{kohout2015device, rabault2016measurements}), or a high accuracy GPS \citep{https://doi.org/10.1029/2019JC015354, KODAIRA2021100567}
to measure the sea ice motion and compute wave spectra. Less common techniques are the use pressure
sensors to measure the wave-induced pressure fluctuation to derive wave observations \citep{marchenko2019wave}. Remote sensing is a promising avenue to obtain observations of waves in ice across large spatial scales \citep{ARDHUIN2017211, horvat2020observing} of far larger extent compared to what can be achieved with in situ instrumentation, though the number of wave parameters that can be retrieved, and the accuracy that can be achieved, is still limited and a topic for open research. Moreover, remote sensing requires considerable more in situ observations for calibration and validation. As a consequence, the most established and accurate method today is
still to deploy instrumentation in situ to measure waves in ice.

Performing in situ measurements of waves in ice is, however, a challenging and
costly operation. Strangely enough, while one would expect access to sea ice to be the
main cost and limitation for such measurements, this is usually not the case in our
experience. Indeed, a quite respectable number of expeditions take place each year into
the Arctic and Antarctic, both scientific, military, and civilian. As a consequence, it has been, in
our experience, relatively easy to piggy back such expeditions and use them as a low cost (from a scientist point of view)
way to deploy instrumentation. By contrast, instrumentation has been a
consistent difficulty point. Commercial instruments have traditionally been extremely
expensive, and sometimes of varying quality, with typical prices after taxes in the 7.5-50kUSD range.
The situation got somehow better with the emergence of new actors, such as Sofar and their
Spotter buoy \citep{smit2021assimilation}, which took down the price to a typical 6-8kUSD cost per instrument, once
import taxes and satellite communication fees are factored in. Still, the Sofar Spotter,
while gaining from being a very modern design, as well as relatively low cost, is not too well
adapted for use in the polar regions, as it was designed primarily for measuring open water
wave statistics, and, correspondingly, relatively large waves. In particular, its battery life limits the duration
of deployments in the polar night \citep{KODAIRA2021100567}. In addition, its GPS-based wave measurement
method, while having many advantages in open water (such as accurate directional wave
spectrum measurement), has a lot higher noise for the typically small waves observed in
the ice than what can be achieved with a modern Inertial Motion Unit (IMU).

\begin{figure}[htp]
  \centering
  \includegraphics[width=.9\textwidth]{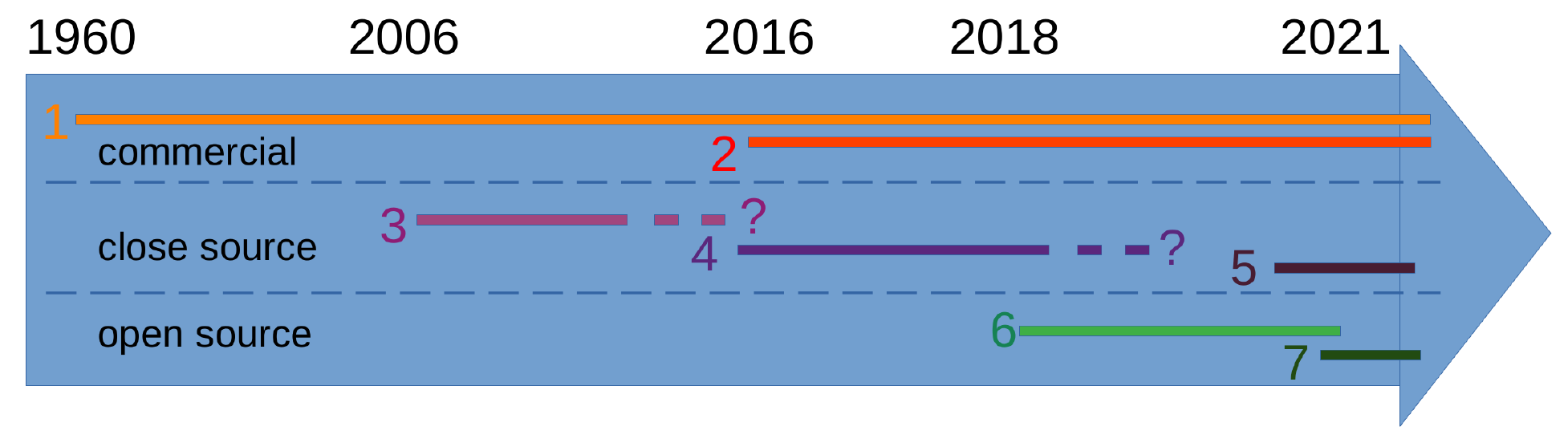}
  \caption{Brief overview of the history of instrumentation for
  measurement of waves in ice using autonomous, satellite- or radio-connected instruments,
  distinguishing between i) commercial instruments (produced and sold by a company), ii)
  close source instruments (research group involved in the design, produced specifically
  for the research group needs), iii) open source (designed by a research group, available
  to all as open source). Commercial instruments for measuring waves have been available
  for a long time, in particular the Datawell Waverider series of instruments ({\bf 1}),
  which goes back all the way to the 1960s \citep{DataWellHistory}. However, such
  "traditional" instruments are often too large and expensive for being a good choice for
    performing measurements in the Marginal Ice Zone (MIZ). More recently, the Sofar Spotter ({\bf 2}) has
  become a viable, lower-cost alternative, though there are a number of limitations in
  particular regarding battery autonomy \citep{raghukumar2019directional}, and the price
  is still significantly higher than what can be attained with open source instruments. A
  number of groups have developed satellite-connected autonomous instruments that have
  been described in research articles, but that remained closed source and which
  blueprints and code were not made available to the scientific community, starting from
  the early 2000s. A few instruments we know about were released in 2006 ({\bf 3})
  \citep{doble2006wave, wilkinson2007autonomous}, 2016 ({\bf 4}) \citep{kohout2015device},
  2021 ({\bf 5}) \citep{johnson2021observing}. By contrast, our group has decided to
  release our designs as open source, so that can be freely used and improved by other
  groups, starting in 2018 ({\bf 6}) \citep{rabault2020open} and now with the present article ({\bf 7}).}
  \label{fig:timeline_general}
\end{figure}

As a consequence of these challenges, a number of groups have developed in-house, custom
instrumentation for measurements of waves in ice
\citep[e.g.,][]{doble2006wave, kohout2015device,rabault2016measurements}. A very brief overview is provided in Fig.
\ref{fig:timeline_general}. This is, in our opinion, a situation that is so far quite inefficient. Indeed,
a number of groups develop their own closed source solution, in parallel of each
other, wasting a large amount of engineering resources in form of working hour and
prototyping costs in the process. This is, in our experience, a major hinder to
1) new groups joining the global waves in ice observation effort, 2) drastically
increasing the volume of observations.

\begin{figure}[htp]
  \centering
  \includegraphics[width=.8\textwidth]{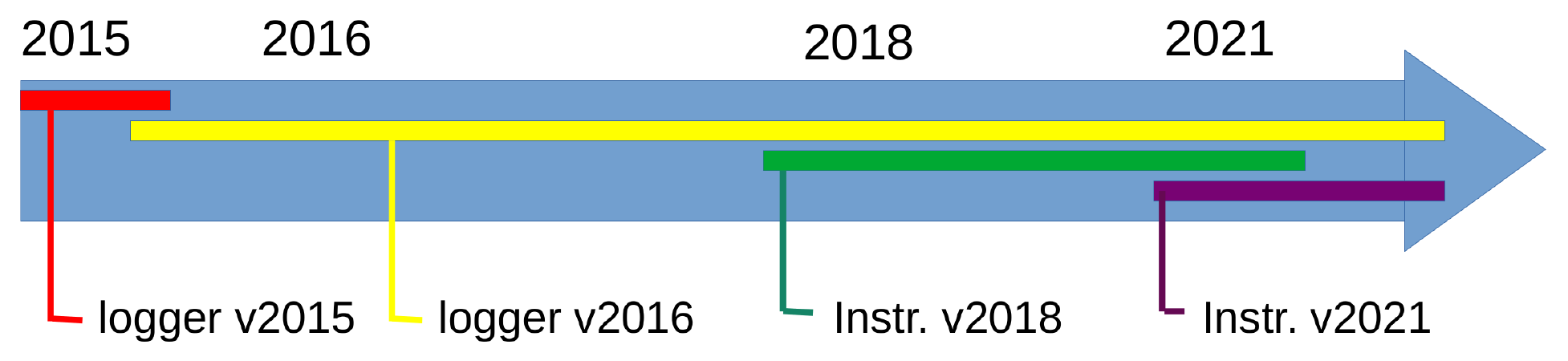}
  \caption{Full lineage of our present open source instrument. Developing reliable autonomous instruments is a long term effort, especially when doing so without resorting to engineering consultancy in order to be able to release the designs open source. The development of our instrument can be tracked back to 2015 \citep{rabault2016measurements} and 2016 \citep{rabault2017measurements}, though the instruments used at the time were simple loggers and did not implement in-situ signal processing nor satellite communication. Using loggers is still an attractive solution when the instruments can be recovered though, and the logger v2016 design is still in use today. One advantage of using loggers when these can be recovered is that the full data timeseries are stored on board on a SD card or similar. By contrast our first truly autonomous, satellite-enabled instrument was designed in 2017-2018 \citep{rabault2020open}, and the present instrument is the second iteration around this design.}
  \label{fig:timeline_our_group}
\end{figure}

An increasingly promising direction is the development of open source
instrumentation, which allows to both reduce hardware costs and mutualize the instrumentation development workload.
This approach was recently adopted by \citet{rabault2020open} and has since its development in 2015 undergone several updates (the full lineage of our series of open source
instruments is presented in Fig. \ref{fig:timeline_our_group}). In the following, we will
refer to the design described in \citet{rabault2020open} as the ``v2018 instrument'', or ``v2018'' in short. Since the initial deployment of the first v2018 instruments \citet{rabault2020open}, the low-cost and open-source fundament of the instrument has promoted usage and international collaboration across the world. The v2018, and adaptations thereof, have been deployed in both the Arctic and Antarctic to study sea-ice drift \citep{sutherland2021determining}, wave attenuation and dispersion \citep{tc-2021-210}, and wave-induced sea ice break-up \citep{voermans2020experimental}. Through these collaborations and technological advancements, points of improvement were identified, which supported to advance the v2018 and develop the latest ``v2021'' version. The aim of this paper is to 1)
publish the design of the v2021 instrument as full open source
software and hardware material, including tutorials, code, post processing codes, and explain why and how the v2021 represents a
major improvement over the v2018, 2) present detailed validation and
performance benchmarking of the v2021 design, 3) offer future prospects to further
federate and drive the efforts in the development of open source instruments and to increase the data set of in situ observations of waves in ice.

These challenges of high instrumentation costs and needs for more in-situ, ground truth data, are not specific to the waves
in ice community. Similarly, the rise of low cost (and possibly open source) electronics is an evolution that is increasingly attracting the attention of both
research groups and public agencies across the world. For example, this trend can be identified also in the
technical development and history of the "Swift drifters" family, among others, which are
also advancing towards open source, low cost, "do it yourself" technical solutions
\citep{thomson2012wave, keating2016fetch, thomson2019new}, focusing on measuring waves
in the open ocean. Another example of this trend is visible in recent works that focus on building small,
inexpensive, bio-degradable surface drifters \citep{smit2021assimilation}. Similarly, the Defense Advanced Research Projects Agency (DARPA) is actively
pursuing similar efforts with its "Ocean of Things" project \citep{darpa_oot}, which aims at
"deploying thousands of low-cost, environmentally friendly, intelligent floats that drift as a distributed sensor network"
(quote from the Ocean of Things project webpage, accessed Dec. 2021). This shows that our developments are of possible interest both
for the waves in ice community, and also, more widely, for many subfields within geoscience that are dependent on gathering large
volumes of field data.

This paper is organized as follows. First, we discuss the general design and features of
the instrument v2021. Then, we present a detailed validation of
the instrument v2021, both i) in a preliminary deployment which aimed at validating the general concept
of the instrument and the power efficiency of the design, ii) in laboratory experiments that tested
the noise threshold of the waves measurements, iii) in a full-scale deployment in the Russian Arctic,
iv) in a full-scale deployment in the Caribbeans. Finally, we offer a few
words of conclusion and we present a roadmap for future development. All hardware
blueprints, embedded software, post processing codes, as well as detailed instructions,
are released as open source on Github, as indicated in the Appendix A.

\section{General design and features of the instrument v2021}

\subsection{Microcontroller}\label{subsec:DesignOverview}

Unlike the instrument v2018, and unlike all other wave measuring instruments that we are
aware of, the v2021 is based solely on a high performance microcontroller for performing
both data acquisition and in-situ data processing. This is a major shift from previous
designs, which usually resorted on some sort of embedded microcomputer for performing the
computation of wave spectra (typical previous solutions, such as those based on a Raspberry Pi, run a stripped down
version of Linux or Windows), with the measurements themselves being
performed either by the same microcomputer, or by a separate, lower power microcontroller. This traditional approach has a dramatic effect on power consumption.
A typical microcomputer consumes around 1 to 2 W (for example, the Raspberry Pi
used in the v2018 consumes typically up to around 350mA while running at a 5V voltage),
while a modern microcontroller can consume as little as 1 to 2 mW (the microcontroller
used in the v2021 consumes typically 500 $\mu$A while running at a 3.3V voltage). This
factor 1000 in power efficiency  allows for significant simplifications and flexibility in the instrument design. This is of particular importance to instrument performance in the polar regions where the battery is generally the main limiting factor in the field experiment duration. In addition, microcomputers will
rely on a complete operating system for running their workloads. These are complex pieces
of software, with several possibilities for things to go wrong, and require the user to
implement a number of techniques to avoid locking of the system. By contrast, microcontrollers can be programmed without the use of any operating system, which allows to build more robust embedded systems that are more reliable in demanding remote environments.

To this end, the instrument v2021 uses an Ambiq Apollo 3 BLU ultra low power microcontroller unit \citep{ambiq}, embedded within a board from the Sparkfun Artemis ecosystem
\citep[Artemis Global Tracker,][]{agt}). This allows to use the complete Sparkfun Artemis
ecosystem toolchain to simplify the programming of the Ambiq Apollo using high level
C++ code. Effectively the user is able to use off-the-shelf, modular, high level libraries and does not need to perform low level drivers and firmware development. The Ambiq Apollo microcontroller used is an ARM
Cortex M4-F implementation that runs at 48MHz (with a 96MHz boost mode), uses 500
$\mu$A of sustained power at 3.3V, and features a complete floating point unit for
performing fast mathematical operations. Since this processor is an ARM Cortex M4
implementation, the full ARM-developed libraries for digital signal processing, including
a wide range of spectral analysis tools, are available out of the box through the Sparkfun
toolchain which includes a copy of the CMSIS library (this features, in particular, real
and complex FFTs, cross correlation computations, and similar, which are available as highly
optimized, ready-to-use functions). The total non volatile memory available for storing
the program is amounting to 1MB, and the RAM has a size of 384kB. To put this in perspective, storing 20
minutes of IMU data recorded at 10Hz into RAM, processing the data, and sending the processed data through Iridium, occupies around one third of the total memory available on the chip. This means that a
lot of additional customized functionality can be added in the future to the v2021.

\subsection{Wave data acquisition and on-board processing}\label{subsec:dataProcess}

The wave data acquisition is performed by a 9 degrees of freedom (9-dof) sensor, which
performs high-frequency measurements of acceleration, angular rates, and magnetic field on
the X, Y, and Z axis of the sensor chip. An industrial quality, thermally compensated, low
noise sensor is used for this (the ST ISM330DHCX together with the LIS3MDL). At
present, only the vertical wave spectrum is computed, and the absolute orientation
(provided by the magnetometer) is not used but can be used in the future to derive directional information. The raw accelerometer and gyroscope data are sampled at approximately 800Hz,
and time averaged into a 100Hz low-noise sensor value. When performing this averaging, a
3-sigma filtering stage is used, i.e., individual measurements that deviate from the rest
of the current 800Hz sample by more than 3 standard deviations are rejected. This allows
to discard occasional bad quality measurements (in our experience, a few such measurements
are obtained in a 20 minutes measurement segment; this is most likely inherent to the kind
of sensor used, and / or may come from rare irregularity in the functioning of
either the 9dof sensor or the firmware used for communicating with it). This low-noise
sensor value averaged at 100Hz is then fed in real time to a Kalman filter run
on the Ambiq Apollo microcontroller at 100Hz. The Kalman filter itself is provided by an
Attitude and Heading Reference System (AHRS) C++ embedded open source library, provided to
the community by Freescale Semiconductors. The Kalman filter
implementation performs data fusion for 9-axis MEMS input (i.e. 3-axis accelerometer,
gyroscope, and magnetometer), and produces an accurate estimate for the absolute orientation
of the sensor in a North, East, Down frame of reference. The Kalman filter output is used
for computing the vertical component of the wave acceleration at 100Hz by projecting the
acceleration from the X, Y, Z frame of reference onto the vertical direction, following
the absolute orientation information provided by the Kalman filter as a quaternion, and
subtracting the constant gravity acceleration. Finally, the processed 100Hz vertical acceleration
measurement is time averaged to 10Hz. The whole procedure is
implemented through a couple of classes in the code and performed instantaneously. The reader
is referred to the open source implementation released
on Github (see Appendix A) for more details.

Resorting to a simple 9-dof sensor and
running the Kalman filter on the Artemis microcontroller is both much more power effective (total power consumption is around 5mA for the proposed solution, compared with typically 35mA using a dedicated full-feature IMU) and significantly less expensive than using a single discrete Inertial Motion Unit
component which performs both tasks on its own (the cost of a high quality 9dof sensor is around 15USD, compared with 50 to 1000USD for a lower accuracy, or similar accuracy, IMU, respectively).
Pseudocode for the data acquisition and Kalman filter processing is presented in the
description of Algorithm \ref{code:kalman}.

\begin{algorithm}
  {\bf initialize} empty BufferVerticalAcceleration10Hz;  \\~\\

  {\bf initialize} KalmanFilter100Hz;  \\~\\

  \While{need more vertical acceleration measurements at 10Hz}{ {\bf initialize} empty
    BufferVerticalAcceleration100Hz  \\~\\

    \For{CurrentVerticalAcceleration100Hz in [0..10[}{ {\bf initialize} empty
      BufferAcc[X,Y,Z]800Hz, BufferGyr[X,Y,Z]800Hz, BufferMag[X,Y,Z]800Hz;  \\~\\

      \For{Current9dofMeasurement800Hz in [0..8[}{ measure Acc[X,Y,Z] and append to
        BufferAcc[X,Y,Z]800Hz;

        measure Gyr[X,Y,Z] and append to BufferGyr[X,Y,Z]800Hz;

        measure Mag[X,Y,Z] and append to BufferMag[X,Y,Z]800Hz; }

      3-sigma-average BufferAcc[X,Y,Z]800Hz into Acc[X,Y,Z]Average100Hz;

      3-sigma-average BufferGyr[X,Y,Z]800Hz into Gyr[X,Y,Z]Average100Hz;

      3-sigma-average BufferMag[X,Y,Z]800Hz into Mag[X,Y,Z]Average100Hz;\\~\\

      {\bf update} KalmanFilter100Hz using Acc[X,Y,Z]Average100Hz, Gyr[X,Y,Z]Average100Hz,
      Mag[X,Y,Z]Average100Hz;  \\~\\

      {\bf obtain} QuaternionAbsoluteOrientation100Hz from KalmanFilter100Hz;  \\~\\

      {\bf compute} CurrentVerticalAcceleration100Hz from
      QuaternionAbsoluteOrientation100Hz, Acc[X,Y,Z]Average100Hz;  \\~\\

      {\bf push} CurrentVerticalAcceleration100Hz to BufferVerticalAcceleration100Hz
      \\~\\
    }

    {\bf average} BufferVerticalAcceleration100Hz into CurrentVerticalAcceleration10Hz;
    \\~\\

    {\bf push} CurrentVerticalAcceleration10Hz to BufferVerticalAcceleration10Hz;  \\~\\
  }

  {\bf return} BufferVerticalAcceleration10Hz; \\~\\

 \caption{Sampling of vertical acceleration at a frequency of 10Hz in a dedicated buffer, including raw data sampling and pre-averaging, Kalman filtering, and vertical acceleration post-averaging. "Acc" stands for "Acceleration", "Gyr" stands for "Gyroscope", "Mag" stands for "Magnetometer". 3-sigma-average is an averaging filter rejecting measurements deviating from the rest of the sample by more than 3 standard deviations, which is used to discard occasional bad readings. The combination of high-frequency averaging of the raw input from 800Hz to 100Hz, Kalman filtering and vertical acceleration computation at 100Hz, and vertical acceleration averaging from 100Hz to 10Hz, allows to balance the need for computational efficiency (by reducing the number of expensive Kalman filter updates) and accuracy (by running the raw data collection at high-frequency and using all the data available through time averaging). This results in a an algorithm that is computationally efficient enough to be run on the Artemis microcontroller in real time, while producing high accuracy measurements.}
 \label{code:kalman}
\end{algorithm}

The wave elevation data processing is also performed on the Ambiq Apollo microcontroller.
At the end of a 20 minutes measurement period (or, to be more exact, 20.48 minutes, to obtain  a multiple of $2048=2^{11}$ samples to allow much faster FFT computation), the
array of 10 Hz vertical acceleration data stored in RAM is processed using the Welch
method. For this, the signal is split into segments of length 2048 sample points, with a
75\% overlap, resulting in a total of 21 segments. The real FFT for the vertical
acceleration of each of these segments is computed using the real FFT implementation
provided by the ARM digital processing reference library, applying Hanning windowing with
an energy-conserving normalization, and these segment FFTs are averaged into the Welch
estimate. The computation of the FFT takes just a few tens of milliseconds due to the dedicated floating point unit on the Ambiq Apollo microcontroller, and the
efficient FFT algorithm provided by ARM. The corresponding algorithm is summarized in
Algorithm \ref{code:waves_proc}. The reader who may want to re-implement a similar algorithm is made aware
that different FFT libraries may use different renormalization conventions, so that using
another FFT implementation may require the use of additional scaling factors. In order to
save memory, the Welch averaging is only performed for the set of frequency bins
corresponding to frequencies that are relevant for waves, which is typically between
$f_{min}=0.05$Hz (i.e. 20s period waves) and $f_{max}=0.5$Hz (ie 2s period waves).

\begin{algorithm}
  {\bf input}: BufferVerticalAcceleration10Hz, length 6 * 2048 samples;   \\~\\

  {\bf initialize} array HanningWelchSpectrum[RelevantIndexRange] with all elements = 0;

  // the RelevantIndexRange includes the reduced span of the 2048-point FFT

  // that covers frequencies between 0.05Hz and 0.5Hz, ie from fmin to fmax   \\~\\

  {\bf initialize} CurrentSignalSegmentStart = 0;

  {\bf initialize} CurrentSignalSegmentEnd = 2048;  \\~\\

  \For{CurrentWelchSegment in [0..21[}{

      {\bf copy}
      BufferVerticalAcceleration10Hz[CurrentSignalSegmentStart..CurrentSignalSegmentEnd[
      into CurrentFFTInput[0..2048[; \\~\\

      // apply Hanning windowing

      \For{CurrentIndex in [0..2048[}{

          CurrentFFTInput[CurrentIndex] *= 1.63 * [sin($\pi$ * CurrentIndex / 2048)]$^{2}$

      }

      {\bf compute} realFFT of  into HanningCurrentFFTOutput;   \\~\\

      HanningWelchSpectrum[RelevantIndexRange] += 2.0 * |
      HanningCurrentFFTOutput[RelevantIndexRange]$^{2} | $ ;

      // the realFFT of a real input signal is a vector of complex numbers; we need to
      extract the

      // complex magnitude squared of each bin; in our case, a renormalization of 2.0 is
      needed since

      // the output is only the half spectrum, i.e. the negative frequencies part of the
      spectrum

      // which is equal to the conjugate of the corresponding positive frequency part is
      omitted \\~\\

      CurrentSignalSegmentStart += 2048 / 4;

      CurrentSignalSegmentEnd += 2048 / 4;
  }
  HanningWelchSpectrum[..] = HanningWelchSpectrum[..] / 21;    \\~\\

  {\bf return} HanningWelchSpectrum; \\~\\

 \caption{Algorithm used for computing the Welch spectrum with energy-conserving Hanning windowing. We have decided to use 21 segments with 75\% overlap when computing the Welch averaging. Note that the exact value of the renormalization coefficients needed may depend on the specific FFT implementation used and the normalization convention that it defaults to. We strongly recommend the reader willing to perform a re-implementation of this processing to carefully test it with synthetic data, so as to avoid renormalization issues.}
 \label{code:waves_proc}
\end{algorithm}

The Welch spectrum with Hanning windowing obtained at this stage is a low-noise estimate for
the spectrum of the wave vertical acceleration, $PSD_{accD}$. From there, the spectrum of
the wave elevation $\eta$ can be obtained following \citep{sutherland2016observations}:

\begin{equation}
PSD_{\eta}(f) = PSD_{accD}(f) / (2.0 \pi f)^4.
\end{equation}

Finally, the spectral moments are computed following:

\begin{equation}
m_i = \int_{f_{min}}^{f_{max}} PSD_{\eta}(f) f^i df,
\end{equation}

\noindent and these are used to compute estimates of the significant wave height $Hs$ and
the wave period $T_z$ and $T_c$, following:

\begin{equation}
Hs = 4 \sqrt{m_0},
\end{equation}

\begin{equation}
T_z = \sqrt{m_2 / m_0},
\end{equation}

\begin{equation}
T_c = \sqrt{m_4 / m_2}.
\end{equation}

We want to remind those readers who may want to re-implement this processing again that
different FFT packets may use different conventions, so that additional renormalization
factors may be needed. Similarly, some references in the literature use the angular
frequency $\omega = 2 \pi f$ rather than the frequency $f$ as a dimension for spectra, in
which case some additional renormalization is needed in the formula above. Our general
recommendation is to test the whole processing algorithm on dummy synthetic data to ensure
that no renormalization factor has been forgotten. Moreover, we used additional checks during the
development of the code to make sure that the scaling and processing as a whole is correct, such as verifying that the Parseval theorem holds by checking that the
variance of the heave displacement is equal to the integral (with respect to frequency) of the power spectral density.

\subsection{Satellite communications}

After the recorded data is processed, the data (composed of the Welch wave acceleration spectrum between $f_{min}$ and $f_{max}$, the estimates for $Hs$, $T_z$, and
$T_c$, as well as UTC timestamp information) are packed into a binary packet. In order to reduce the size of the binary
packet, the estimates for $Hs$, $T_z$, and $T_c$ are transmitted as 32-bit floats, while
the Welch frequency bins are discretized into renormalized 16-bit unsigned integers. The
renormalization of the Welch bins is performed relatively to the peak value of the
spectrum, which is transmitted as a 32-bit float within the packet. This allows to
significantly reduce the size of the binary packets and the overall iridium costs while
keeping a high accuracy for the data transmitted.

In addition to the wave spectrum, geographical positioning is obtained with a simple GNSS
module, and is also used to generate accurate UTC reference times and to
periodically re-calibrate the real time clock of the microcontroller to avoid time drift.
The GNSS data are packed, buffered, and transmitted using an efficient binary encoding, similar to what is done
for the wave data.

The sample rate of both the GNSS and the wave measurements can be adapted through 2-way
iridium communication. In addition, the real time clock present on board on the microcontroller is used to make sure that the measurements are performed at fixed hours and minutes. In order to reduce
iridium costs and energy consumption, the firmware can pack several binary packets
packages together before transmitting these as a single iridium message. As all
iridium communications are buffered on the microcontroller, any data not transmitted due to failure in iridium communication can be retransmitted at the next transmission attempt to prevent data loss. When an Iridium communication is established, the instrument transmits a burst
of several messages, sending back to the user all the information that are stored in the
binary message buffers.

A simple binary protocol decoder, available as a Python module, is provided alongside the
code for the instrument firmware. This, together with the web interface or web API offered
by modern Iridium providers (in our case, we use Rock Seven Mobile Services Ltd, though
other providers would be possible), allows great flexibility and ease-of-use for the end
user. Those tasks can easily be automated, and we provide a custom bash script for performing
https requests directly to the server of RockSeven, which allows to retrieve all the iridium
messages received over a user-selected time span, as a csv database.

\subsection{Ongoing instrument variant: cellular communication}%

A derivate of the satellite based buoy is being developed to communicate
through the cellular network instead. Using the same data-processing and setup means
that the buoy takes advantage of the previous validation. The purpose of a buoy
operating on the cellular network is 1) lower cost (by a factor of 2 to 3 for the hardware,
and up to 50 for the communication), 2) higher
data-rates: time-series of continuous measurements can be transmitted if necessary.

The buoy will operate under the limitation that it cannot stray far from the
coast, and the target areas reflect this: near-shore breaking wave measurements
and drift and wave measurements inside fjords in Norway. The buoy will be
designed to store as much data as possible locally if network service is
temporarily unavailable. If the buoy runs out of memory in the mean time it
will drop uncritical data, and store only processed data. In addition to the
internal memory about 900kB of additional memory is available through the
cellular modem for caching data, and other memory devices may be added at small
extra cost and power usage if necessary.

\subsection{Battery autonomy and power-saving strategies}

Reliable power supply is a critical concern for waves in ice instruments due to the cold temperature conditions and the inability for solar power usage during polar night. In the present design, we decided to avoid the use of solar panels
altogether (though solar panels could be easily added for building an open ocean buoy with unlimited autonomy outside of the polar regions). This allows to simplify the design, makes it less expensive and less labor
intensive to produce, as well as avoids compromising the water-tightening of the instrument. In our case, the combination of a power-efficient microcontroller and 9dof, as well as power-optimized firmware, led to a drastic reduction of
the power consumption compared with previous generations of the instrument. Table
\ref{table:power_use} summarizes the power consumption of the instrument in different
modes. We use as power sources Lithium D-cells (SAFT LSH20), that have a nominal capacity
of around 13Ah at 3.6V. In addition, a 3.3V step up-step down buck converter
\citep{pololu} is used to provide a stable 3.3V power source, even when the iridium modem briefly
needs to draw much more power (bursts up to 250mA). This results in a typical operational
time, with 2 Lithium D-cells mounted in parallel and no solar power, of around 4.5 months with
our standard GNSS and waves measurement rates (GNSS position every 30 minutes and wave spectrum every 2 hours).
A solar panel could easily be added to the design for long term deployments outside of the polar regions,
in which case even a very small solar panel (for example, a 1W solar panel, which costs typically 20USD and has a size
of 10cm x 10cm), coupled to a small rechargeable battery, would be more than enough to provide the
18.7mW of average power needed to sustain operation.
Otherwise, if needed, increasing the number of batteries used in parallel will allow to increase the autonomy,
proportionally to the number of cells used. In order to simplify the handling and deployment of the instrument, a magnetic switch was added to the design \citep{reed_switch}. When the
magnet is attached, the instrument is powered down. The
instrument starts operating as soon as the magnet is removed, and can be switched on and
off as many often as necessary. Some LEDs are used to provide visual indication
of instrument activity, while their blinking frequency is taken low enough to preserve battery.

\begin{table}[h]
\begin{center}
\begin{tabular}{|| c | c | c | c | c ||}
  \hline
activity mode & activation frequency  & current (mA)  & mWh use per hour & time to empty 2
Li D-cells \\
\hline
\hline
sleep & when not active & 0.3 & 1.0 & 7.3 years \\
\hline
gnss measurement & 2 minutes twice per hour & 30 & 3.3 & 2.2 years \\
\hline
wave measurement & 20 minutes every 2 hours & 8 & 4.4 & 1.6 years \\
\hline
iridium transmission & 1 message per hour & burst 250mA $^{*1}$ & 10 $^{*2}$ & 0.7 years
\\
\hline
\hline
typical use & & & 18.7 & 0.39 years $\approx$ 4.6 months \\
\hline
\end{tabular}
\\~\\
\caption{Overview of the power consumption and resulting battery autonomy resulting from
the different components of the instrument when using two non rechargeable D-sized Lithium
batteries. 'Activity mode' indicates which part of the workflow of the instrument that is
being considered. 'Activation frequency' indicates a typical activation duration and
frequency (though this is only indicative, and may be modified either by using different
parameters when programming the instrument, and / or by sending updated activity rates
through the bi-directional iridium communication). 'Current (mA)' is the average current
used in the corresponding mode when it is active. 'mWh use per hour' indicates the
corresponding energy use for the given activation mode, following the formula:
$\text{mWh} \text{UsePerHour} = \text{Current(mA)} \times 3.3\text{(V)} \times \text{ActivationFraction}$. The 'time to
empty 2 Li D-cells' (more specifically, 2 LSH20 batteries), is computed using the following assumptions: individual LSH20
battery voltage: 3.6V during the full discharge cycle, and capacity: 10Ah (which is a
pessimistic estimate to take into accounts the effect of low temperatures, as datasheet
value is 13Ah), with voltage converter efficiency of 90\%, i.e. total energy available
with 2 cells: 64Ah. The time to empty in years is then equal to $64e3 / mWhUsePerHour / 24
/ 365$.  All values reported here are based on measurements in the laboratory with the
carefully power-optimized firmware, and with "always on" LEDs de-activated by cutting the
traces supplying power to these, following the instructions manual of each of the
individual breakout boards used. Footnotes: $^{*1}$: the Iridium modem uses high transient
currents; a couple of super capacitors are already present on the Artemis Global Tracker
main board to smooth these out. $^{*2}$: since i) there are high transient currents when
transmitting a message that are challenging to measure accurately, and ii) total energy
used may depend on the quality of the signal, this quantity is quite difficult to
estimate. In the present case, we use the value provided by the manufacturer \citep{iridium_power}; this seems
to be a pessimistic estimate, see the next section of the first deployment test.}
\label{table:power_use}
\end{center}
\end{table}

The high level logics of the instrument (wake-up patterns, measurements patterns, iridium
burst mode transmission) is implemented as a few lines of code that leverage an object
oriented implementation of the processing components described above. To increase robustness of the firmware after instrument malfunction, we resort
to a double strategy. First, the firmware is coded following defensive programming
patterns, with extensive quality checks of all inputs and outputs for the different
modules used. Second, the integrated hardware watchdog present on the microcontroller is
enabled at all time, in order to force a complete reboot of the microcontroller in case
of a hardware or a software malfunction. The test deployments have all been highly successful
as described in Section 3, which is a testimony to the robustness of the design.

\subsection{Total cost and assembly process}

The total bill of materials is presented in Table \ref{table:components}. We consider that around 0.5 hours of work is needed to assemble one single instrument. No advanced
electronics or hardware experience is required to build the instrument. At present, the
assembly relies on a couple of soldering steps to connect the power supply module to the
main board, and simply plugging the 9dof sensor and the electronic switch controlling it
into the main board using a couple of Qwiic I2C cables.

\begin{table}[h]
\begin{center}
\begin{tabular}{|| c | c | c | c ||}
  \hline
component & function  & price (USD)  & assembly steps \\
  \hline
\hline
Artemis Global Tracker & main board, MCU, GNSS, Iridium & 375 & ready to use  \\
\hline
GNSS + Iridium antenna & passive antenna & 65 & screw on SMA cable  \\
\hline
SMA extension cable 25 cm & extension cable for antenna & 5 & screw on tracker  \\
\hline
Qwiic power switch & power on and off 9dof & 7 & disable LED, connect 9dof and tracker  \\
\hline
ISM330DHCX + LIS3MDL & 9dof sensor & 18 & connect to power switch \\
\hline
Qwiic cables (x2) & connect tracker, 9dof, switch & 3 & connect power switch and 9dof \\
\hline
3.3V Regulator S7V8F3 & 3.3V buck converter & 10 & solder to battery and tracker \\
\hline
2 x D cell holders & house and connect batteries & 15 & solder to 3.3V regulator \\
\hline
2 x SAFT LSH20 & power supply & 35 & put in cell holders \\
\hline
reed MDRR-DT-20-35-F& magnetic switch & 3 & solder between battery and regulator \\
\hline
magnet & turn magnetic switch on / off & 1 & mount outside housing \\
\hline
housing box & housing, IP68 & 20 & mount the electronics inside \\
\hline
misc: glue, wire & small extras & 5 & get the design assembled \\
\hline
\hline
total & fully functional instrument & 562 & 0.5 hours / instruments, producing 10 \\
\hline
\end{tabular}
\\~\\
\caption{A representative list of components needed to build an instrument monitoring drift (GNSS) and wave activity (9dof sensor). Additional components will be necessary to extend functionality with features such as pressure and temperature measurements. All prices are estimates gathered in October 2021 and may vary with time. In addition, some delivery costs and import fees may apply. In Norway, ordering components in bulk typically results in a total price per instrument of around 650USD. The assembly time for a single instrument, when assembling a series of 10 instruments in bulk, is about 0.5 hours once the user is familiar with the design. For more information about the components, where these can be ordered, programmed, and assembled into an instrument, see the Github repository (more information in Appendix A) and the assembly manuals and instructions there.}
\label{table:components}
\end{center}
\end{table}

\begin{table}[h]
\begin{center}
\begin{tabular}{|| c | c | c | c ||}
  \hline
functionality & credits / message  & messages / hour (default)  & price / month (USD) \\
  \hline
\hline
iridium subscription fee 1 month & N/A & N/A & 16 \\
\hline
GNSS position data & 2 & 0.3 & 26 \\
\hline
wave spectrum data & 3 & 0.5 & 66 \\
\hline
\hline
total & N/A & N/A & 108 \\
\hline
\end{tabular}
\\~\\
\caption{Overview of the Iridium communication costs. The prices correspond to the ones offered by RockSeven at the time of writing this manuscript, prices may change with time and depending on conversion rates between currencies. One Iridium communication credit (corresponding to 50 bytes of transmitted data) costs 0.06 USD in the biggest credit bundle. In addition, each individual modem costs a flat price of around 16 USD to be activated per month. For a default-configurated instrument, which measures GNSS position each 30 minutes, and wave spectrum each 2 hours, the monthly Iridium communication costs are computed following the formula: $\text{PricePerMonth} = 30.5~\text{DaysPerMonth} \times 24~\text{HoursPerDay} \times \text{NumberMessagesPerHour} \times \text{NumberCreditsPerMessage}$ . Of course, it is possible to modify the rate of data collection by either using different setups when one programs the instrument firmware, or by sending a measurement rate update command to the instrument through iridium. This will reduce prices accordingly (except for the subscription fee part), for example, measuring the GNSS position once an hour and measuring wave spectrum once each 4 hours would bring down costs to 62 USD per month.}
\label{table:iridium_costs}
\end{center}
\end{table}

The total weight of the instrument depends on the size of the enclosure and the number of
battery cells used. Typical weight is between 0.3 and 0.5kg with 2 lithium D-cell batteries. The
dimensions of the instrument are typically around 12cm x 12cm x 9cm. Many of these
parameters are part of design tradeoffs. For example, more batteries increase weight and
would make deployment from drones more challenging, but also allow to extend battery time.
The use of a small enclosure allows for easier shipping and logistics and make the
instrument less likely to be discovered (and destroyed) by polar bears (in the Arctic), but means that
even a relatively minor snowfall is enough to bury the instrument and interrupt GNSS and
Iridium signals.

In addition to the construction costs, Iridium costs can be significant over time. Table
\ref{table:iridium_costs} indicates the typical Iridium communication costs per month, for
an instrument running with the default parameters (reducing the frequency of measurements
will reduce costs accordingly).

Combining the building and the Iridium costs, the total cost per instrument, for an
activity period of 6 months (which is reasonable for instruments deployed in the MIZ at
the start of the ice formation season), is around 1200 USD, all included (noting that 4, rather than 2, LSH20 batteries will be necessary for such a deployment).
Around 53\% of this cost comes from the Iridium transmission fees, so reducing the
frequency of measurements when little activity of interest is happening will allow to
significantly reduce the total cost of ownership.

\section{Validation of the instrument v2021}

\subsection{Autonomy and satellite communication test in the Arctic: February 2021 deployment}

An early version of the instrument v2021, which performed drift tracking but no waves
measurements, was deployed East of Svalbard in February 2021 \citep{nilsen2021pc}. This early version of the
instrument was used to i) validate the general design of the electronics, ii) validate the
low power modes and the ability of the instrument to live for several months in the Arctic
cold using just a couple of D-sized battery cells, in particular, regarding the power
consumption of the iridium modem which is hard to estimate, iii) validate the satellite
communication binary protocols.

This initial test was a success. The instrument was
deployed on February 24th, 2021, East of Svalbard. It then transmitted data for a few
weeks, before a snow storm covered it with snow and blocked the iridium transmission. This
loss of communications is a consequence of the design of the housing of the instrument,
that relied on a box that was too low to survive even just 10 cm of snow, and will be
addressed in future designs, by using a box that is higher and does not get buried as
easily (though there is a tradeoff in this design choice, as a larger box is also more likely
to attract polar bear attention and get destroyed). However, communication with the Iridium satellites was restored when the ice
melted in late May 2021. Following this, the instrument freely floated in the open sea,
traveling all the way from Svalbard to the close vicinity of Murmansk, and back and forth between
Novaya Zemlya and Murmansk.

\begin{figure}[htp]
  \centering
  \includegraphics[width=.7\textwidth]{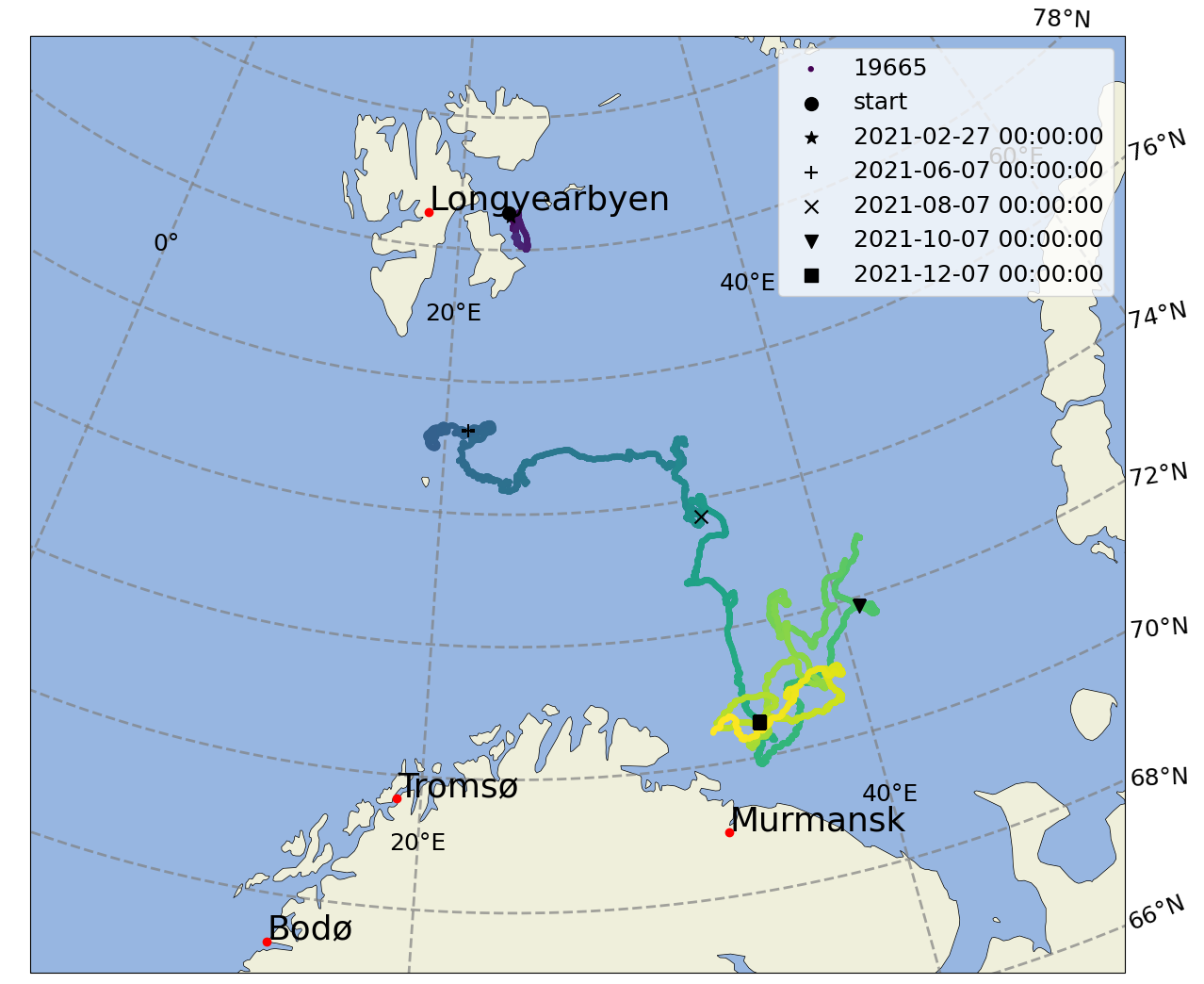}
  \caption{Overview of the drift trajectory of the early prototype instrument deployed in February 2021 East of Svalbard. The interruption in the trajectory is due to a layer of snow covering the instrument that prevented satellite communications during a couple of months. The design will be modified to use an enclosure that is higher and does not get buried as easily. The instrument then drifted freely in the open ocean. This validates the low power design, and confirms that the design chosen is able to sustain operations for several months using just a few D-cell Lithium batteries.}
  \label{fig:feb2021_deployment}
\end{figure}

An overview of the trajectory is presented in Fig. \ref{fig:feb2021_deployment}. At the time of
writing this manuscript, i.e. January 2022, this instrument is still transmitting.
The instrument was powered by 3 LSH20 batteries, which, since it
measures only GNSS position each 30 minutes, should result in a battery life of around a
bit over 1 year following the data from Table \ref{table:power_use}. Therefore, this
deployment validates the low power design, and confirms that both the low power sleep
mode, the GNSS data acquisition, and the Iridium transmission (which have the power
consumption that is most difficult to estimate), are correctly implemented and allow
extended operation on battery alone.

\subsection{Wave tank experiments for validation of small-amplitude, low frequency wave measurements
at the University of Tokyo's wave-ice tank facility: July 2021 laboratory test}

\begin{figure}[htp]
	\centering
	\includegraphics[width=.7\textwidth]{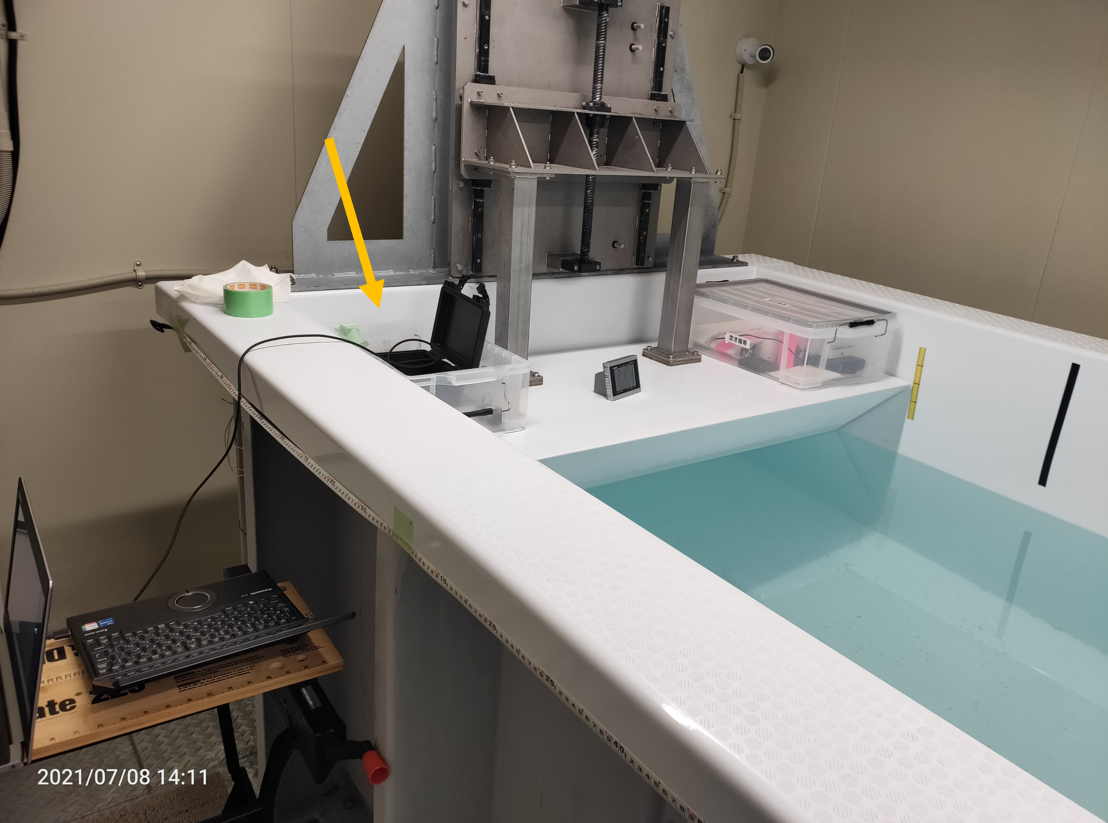}
	\caption{Illustration of the wave tank experiment setup in the wave-ice tank of JAMSTEC (Japan Agency for Marine-Earth Science and Technology) at the Kashiwa campus of the University of Tokyo, Japan. The instrument v2021 location is shown by the yellow arrow.}
	\label{fig:tankImage}
\end{figure}

We conducted a wave tank experiment in July 2021 to evaluate the instrument v2021 accuracy
for small amplitude waves. This experiment allows to fully validate i) the acquisition of
raw data from the 9dof sensor, ii) the processing algorithm and scaling factors used in
the low-level signal processing code, iii) the data encoding used for transmitting the
information back to the user. The experiment took place in the wave-ice tank facility at
the Kashiwa campus of the University of Tokyo, Japan. The wave tank is 8 m long and 1 m
wide with 0.6 m water depth. Waves are generated by a wedge-shaped plunger-type
wavemaker, and there is an artificial beach at the other end to minimize wave reflection.
In our case, however, the instrument v2021 prototype was placed on the top of the
wavemaker rather than in the water (see Fig. \ref{fig:tankImage}), so it measured the
vertical motion of the wavemaker, which is known to a very high accuracy thanks to the use
of a close-loop control.

The experiment aims to evaluate the instrument accuracy when it measures centimeter scale wave
signals at low frequencies corresponding to waves in ice conditions. As such, we tested
the instrument using both 1 and 2 cm amplitude monochromatic waves for 12, 14, and 16 s
wave periods (as well as an "extreme" test with 0.5 cm amplitude and 16 s period, to test
the very limit of the instrument sensitivity), which resemble lower-bound wave signals that were previously measured in
ice-covered ocean \citep{voermans2020experimental}. The spectral analysis setup was slightly modified from that of
Section~\ref{subsec:dataProcess}, so that each test case duration was reduced from 20 mins
to roughly 7 mins, in order to be able to perform the tests within the allocated time slot.
For this, the segment overlap was reduced to 50\% (corresponding to 1024 points),
which yields a total number of 3 segments per 7 mins measurement interval. All other
processing configuration remained the same as Section~\ref{subsec:dataProcess}.

The wave tank experiment results are summarized in Table \ref{tab:wavetankSummary}. The
frequency of the wavemaker motion was not tuned to the exact frequencies bins,
nevertheless, the peak of the energy was consistently captured in the nearest frequency
bin, as reflected in Table \ref{tab:wavetankSummary}. Then, the best estimate of the wave
amplitude can be derived as $A_{mono} = \sqrt{df \times \left( S(i-1)^{2}+S(i)^{2}+S(i+1)^{2} \right) }$,
where $S(i)$ is the wave spectrum taken at the peak of the amplitude spectrum,
and we consider also the neighboring frequency bins. The results of the wave tank experiments
summarized in Table \ref{tab:wavetankSummary} indicate
that the instrument is capable of measuring monochromatic wave amplitudes of 1 to 2 cm
with roughly 0.1 cm accuracy.

\begin{table}[!htbp]
	\centering
	\begin{tabular}{|l|l|l|l|l|}
		\hline
		\begin{tabular}[c]{@{}l@{}}Wavemaker signal: \\ amplitude, period\end{tabular} &
		\begin{tabular}[c]{@{}l@{}}Reported buoy amplitude \\ (cm, averaged)\end{tabular} &
		\begin{tabular}[c]{@{}l@{}}Reported buoy \\ peak period (s)\end{tabular} &
		\begin{tabular}[c]{@{}l@{}}frequency bin closest \\ to peak frequency\end{tabular} &
		Number of cases \\ \hline
		2 cm, 16 s                                                                    & 2.104
		& 15.75    & yes                & 4               \\ \hline
		1 cm, 16 s                                                                    & 1.040
		& 15.75    & yes                & 2               \\ \hline
		0.5 cm, 16 s                                                                    & 0.62
		& 15.75    & yes                & 2               \\ \hline
		2 cm, 14 s                                                                    & 1.954
		& 13.65    & yes                & 1               \\ \hline
		1 cm, 14 s                                                                    & 1.053
		& 13.65    & yes                & 1               \\ \hline
		2 cm, 12 s                                                                    & 2.022
		& 12.05    & yes                & 2               \\ \hline
		1 cm, 12 s                                                                    & 0.970
		& 12.05    & yes                & 2               \\ \hline
	\end{tabular}
	\caption{A summary of wave tank experiment results. The experiments included 1 and 2 cm amplitude waves for 12, 14, and 16 s periods, as well as an "extreme" test corresponding to an amplitude of 0.5cm at 16s period, to test the very limit of the instrument sensitivity. When multiple tests were conducted, the mean values are reported. We find that the instrument is able to accurately measure waves down to 1 cm at a 16s period, with a corresponding typically accuracy of under 0.1 cm. For waves of amplitude 0.5cm and period 16s, the peak period is still correctly identified, but noise in the measurements starts to play a role, with a deviation of around 25\% in the wave amplitude value reported.}
	\label{tab:wavetankSummary}
\end{table}

\begin{figure}[htp]
	\centering
	\includegraphics[width=.5\textwidth]{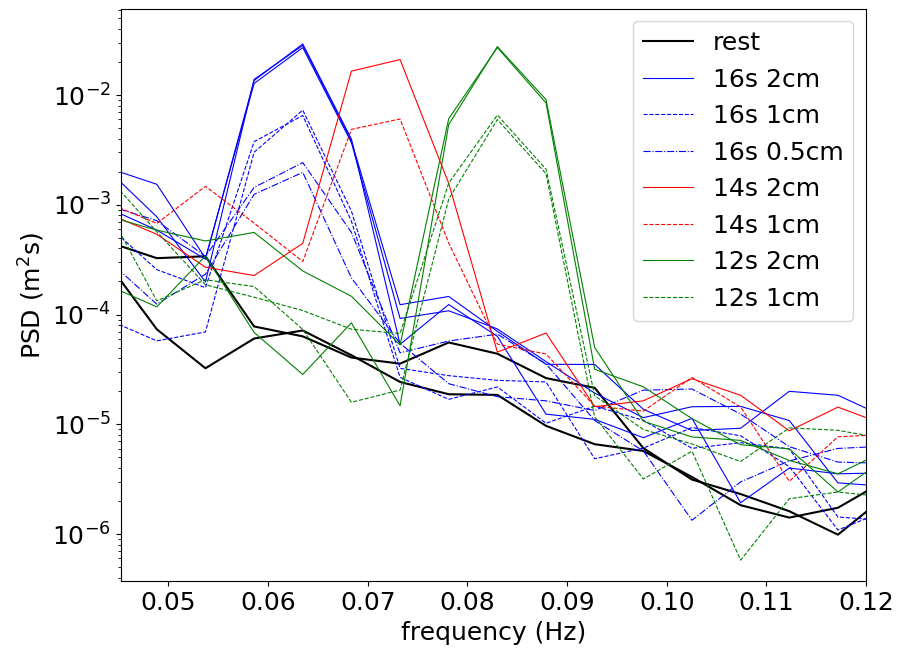}
	\caption{Spectra obtained when testing the instrument v2021 in the laboratory. This confirms the excellent
  accuracy of the instrument, which was already reported in Table \ref{tab:wavetankSummary}. In particular,
  the signal-to-noise ratio is superior or equal to 50 for all test cases, except the "extreme" test case with
  displacements of amplitude 0.5cm and period 16s, where the signal-to-noise ratio is around 10. The noise threshold
  obtained when the instrument is at rest is actually slightly better than what was obtained with the instrument
  v2018 (compare the present plot to Fig. 3 of \citet{rabault2020open}, curves "IMU 1" and "IMU 2"). This is an impressive result,
  given that the instrument v2018 used a Vectornav V100 IMU which costs around 1400USD all included, while the present
  instrument v2020 uses a 9-dof sensor that costs around 20USD all included.}
	\label{fig:lab_spectra}
\end{figure}

The spectra obtained during these tests, as well as two spectra corresponding to the instruments
being at rest, are presented in Fig. \ref{fig:lab_spectra}. As visible there, the noise background
of the instrument increases slightly when motion is present, which may be explained by a variety of
reasons (small vibrations in the wave paddle mechanism when it is undergoing displacement, diffusion of
numerical noise in the processing algorithms, inherent properties of the 9dof sensor). However, the
increase in the noise threshold when there is motion remains very limited. The signal-to-noise ratio
is superior or equal to a factor of 50 in all experiments, except for the smaller waves (0.5cm at 16s period),
in which case the signal-to-noise ratio is around 10. This means that, even in this last, "extreme" test case,
the signal-to-noise ratio is still good enough to clearly distinguish the motion from the noise floor,
though noise starts to be visible in the integrated statistics, as discussed in the previous paragraph and
in Table \ref{tab:wavetankSummary}.

Following this successful laboratory validation, we decided to perform a field validation
experiment in the Arctic, which is described in the next subsection.

\subsection{2021 NABOS expedition and comparison with SOFAR Spotter buoy in the Marginal Ice Zone in the Arctic: September 2021 deployment}\label{subsec:NABOSdeployment}

The first field deployment opportunity for the full-feature
instrument v2021 was the 2021 Nansen and Amundsen Basins Observational System (NABOS)
expedition (\url{https://uaf-iarc.org/nabos-cruises/}), which was performed in the context of the ArCS II
Japan-Russia-Canada International Exchange Program
(\url{https://sites.google.com/edu.k.u-tokyo.ac.jp/arcsii-iep-jrc/main}).

A v2021 instrument was assembled in a Zeni Lite drifting buoy, referred to Zeni-v2021
in the following (see Appendix B for the buoy assembly details). This instrument was deployed alongside
a commercial wave measuring device (SOFAR Spotter drifting buoy,
\url{https://www.sofarocean.com/products/spotter}), referred to as SPOT-1386 in the following. The
key difference of the Spotter compared with the instrument v2021 is that the Spotter measures surface
displacements based on GNSS signal.

\begin{figure}[htp]
	\centering
	\includegraphics[width=.9\textwidth]{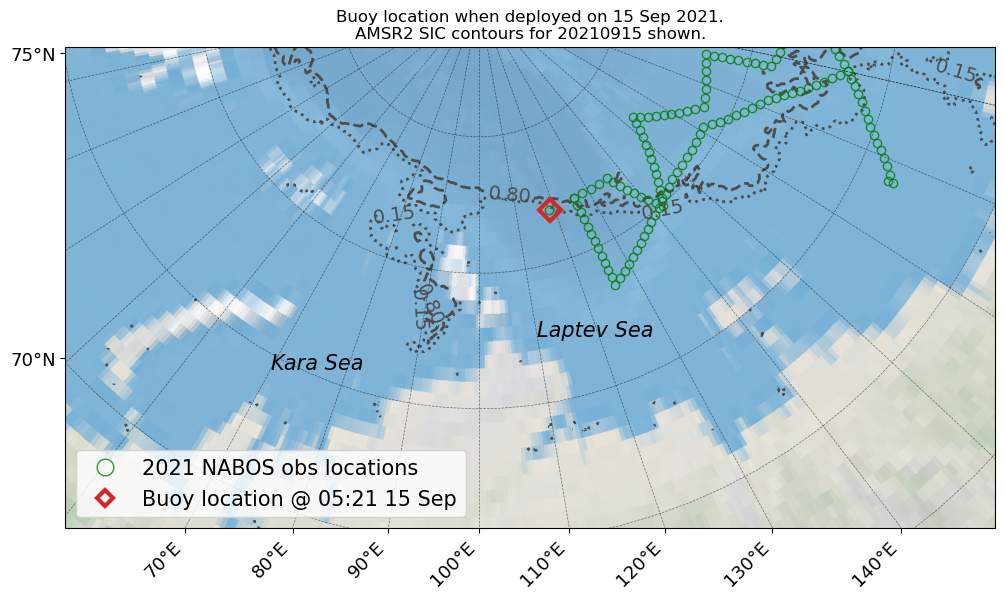}
	\caption{Zeni-v2021 and SPOT-1386 deployment location on 15 Sep 2021. The 2021 NABOS observation locations also shown in green markers, and AMSR2-derived 0.15 and 0.80 Sea Ice Concentration (SIC) contours for the same day are overlaid.}
	\label{fig:ZeniSpotMap}
\end{figure}

On 15 September 2021, both buoys were deployed adjacent to the ice edge in the central Arctic
Ocean (North of the Laptev Sea), at a location of 81.915$^\circ$ N 118.763$^\circ$ E, at around 05:05 UTC.
The buoy deployment location is shown in Fig
\ref{fig:ZeniSpotMap}, which is overlaid with 0.15 and 0.80 sea ice concentration (SIC)
contours based on the Advanced Microwave Scanning Radiometer 2 (AMSR2) data
\citep{ADS-AMSR2}.

\begin{figure}[htp]
	\centering
	\includegraphics[width=.45\textwidth]{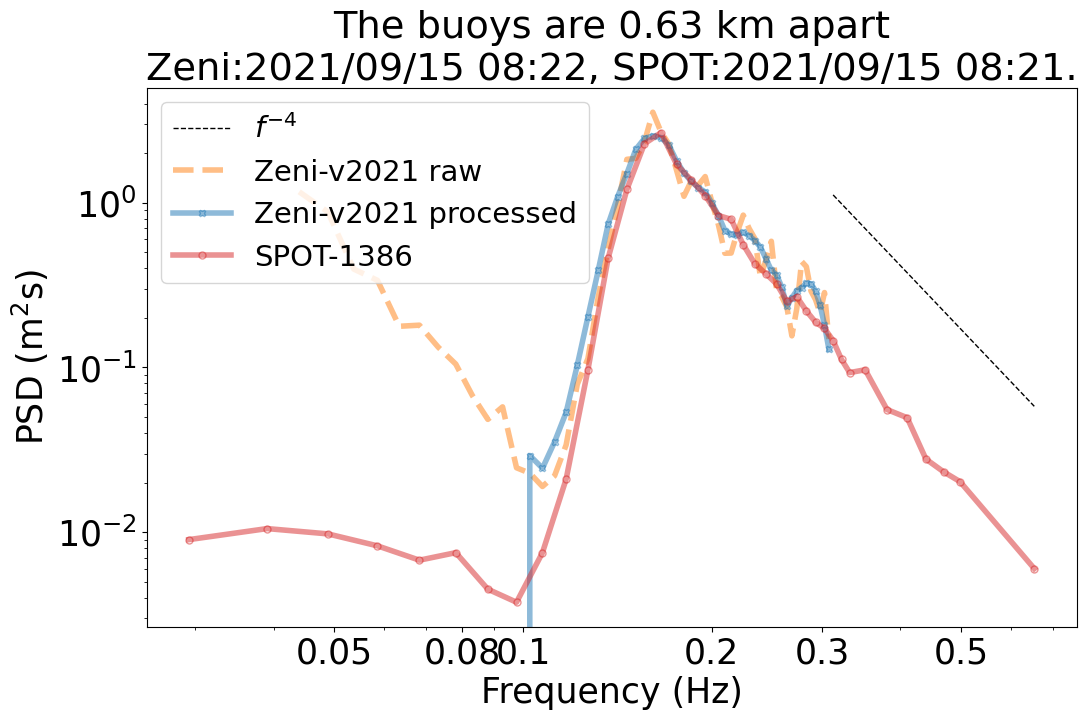}
	\includegraphics[width=.45\textwidth]{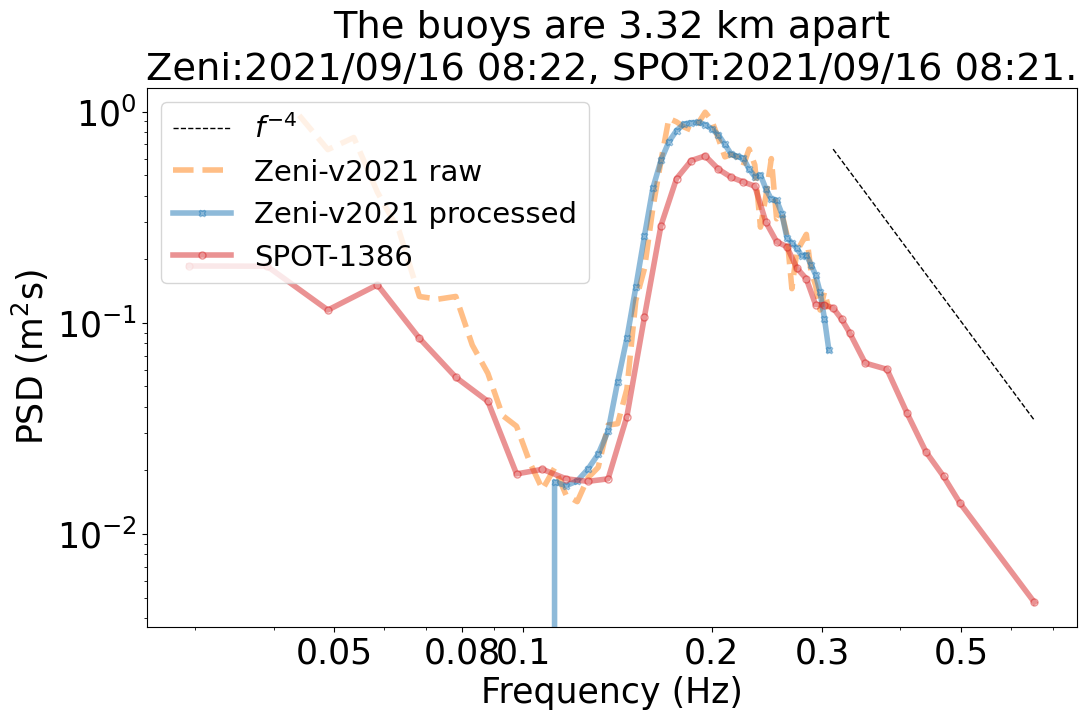}
	\caption{Comparison of Zeni-v2021 and SPOT-1386 spectra at 08:21 15 Sep 2021 (left) and 08:21 16 Sep 2021 (right): the distance between the buoys were roughly 1 km and 3 km, respectively. Low frequency noise of Zeni-v2021 spectra was removed by applying an ideal filter at the lowest frequency local minimum of the smoothed spectrum,
  similarly to what was reported in \citet{Waseda2017, Nose2018}, which is a common post processing step when working when accelerometer-derived wave data. This indicates excellent agreement between the Sofar Spotter and the instrument v2021 measurements in the open water.}
	\label{fig:v2021Spectra}
\end{figure}

The two buoys drifted apart roughly 1 km within 8 hours of being deployed and farther than 3 km
after 24 hours. Given that the ocean surface conditions near the ice edge are heterogeneous, we
compared the instrument v2021 spectrum with that of SPOT-1386 following the deployment, when they are
still close to each other. The buoy spectra compared generally well when the buoy distance was
less than 1 km. The left panel in Fig~\ref{fig:v2021Spectra} shows the buoy spectra at
08:21 15 Sep 2021, which corresponds to a moderate significant wave height $H_{m0}$=1.62~m. The comparison
indicates that the spectra compare well when the low frequency noise, which is typical in
accelerometers, is discarded. Here, the method that was previously used in
\cite{Waseda2017,Nose2018} was applied to remove this noise from Zeni-v2021 spectra. An
ideal filter was applied to remove data below the lowest frequency local minimum of the
smoothed spectrum. In addition, another spectra comparison on 08:21 16 Sep is also
provided on the right panel in Fig~\ref{fig:v2021Spectra}. By this time, the buoys were
farther than 3 km apart, so there is a slight difference in energy levels between the two buoys, but the general shape of the spectra is comparable. These buoy
spectra comparisons indicate that the instrument v2021 can be used to measure
ocean waves, and is a full scale, fieldwork validation of the algorithms and processing used.

Spotter buoys, without solar radiation recharge at high latitude during the winter season, have only a short life span and SPOT-1386 ran out of
battery on September 30th 2021. Analysis into the co-located deployment in and near the Marginal Ice
Zone MIZ is ongoing and will be presented in a separate, science-focused paper. Notwithstanding, a comparison of wave periods (peak period $T_p$ and
spectral period $T_{0m1}$) when the buoys were less than 5 km apart show very good agreement (not shown here, as the data have scientific value that goes beyond validation
and are used in an ongoing scientific work). In addition, the part of the dataset for which both
instruments are frozen in the ice indicates that the noise level of the IMU-based instrument (typically around 0.1cm) is much better
than that of the Spotter buoy (typically around 1cm), which is expected due to the different data sampling techniques used (GPS vs. 9dof). This
provides further support that Zeni-v2021 captured consistent energy densities to those of
SPOT-1386, and that the Zeni-v2021 is actually better suited for measuring small waves in ice. Note that the analysis into the
Zeni-v2021 and SPOT-1386 co-located deployment is described in a manuscript under
preparation, which is why these data are not released in the present technical paper.

Regarding battery life, the estimated deployment duration for hourly wave sampling rate
was 3.3 months based on Table \ref{table:power_use} where 2 Tadiran D-cell batteries
(TL-5930/S), each with 19 Ah capacity, were used as the power source. The instrument stopped transmitting, likely
due to empty battery, in mid-December, i.e. 3 months after the start of the deployment. This corresponds well
to the estimated battery life, and this is an additional
confirmation of the quality of the design of the hardware and software, in particular regarding the implementation of low
power modes.

\subsection{The "Floatenstein" drifting buoy in the Caribbeans: November 2021}

\begin{figure}[htp]
	\centering
	\includegraphics[width=.45\textwidth]{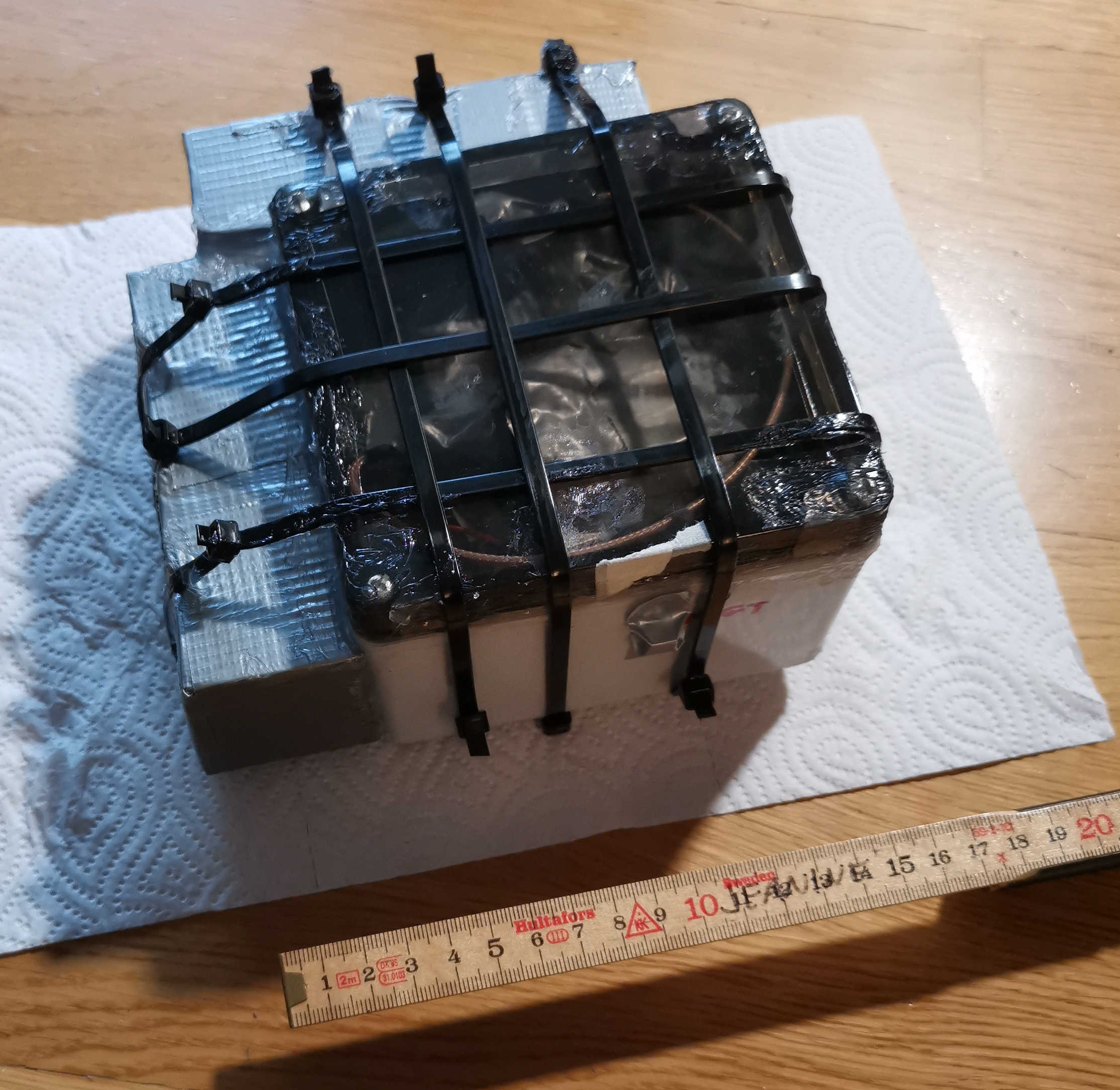}
	\caption{The "Floatenstein" drifting buoy. The main body of the buoy is a rectangular plastic box of dimensions 12cm x 12cm x 9cm. Due to the use of alkaline batteries
instead of lithium batteries, additional floatation elements were needed. These are the two rectangular elements on the side of the main body of the buoy, which are built
from styrofoam chunks, wrapped in duct tape, and held in place by cable ties. The full box is sealed with epoxy to be perfectly water tight. To avoid the need for making holes in the box, the Iridium antenna is also taped in place inside the box, immediately under the lid, pointing 45 degrees upwards. The small cubic magnet (wrapped
in duct tape to hold it in position under transport), which controls the magnetic switch glued on the inside of the box, is visible on the front panel of the main body of
the drifter. Despite its unusual appearance, Floatenstein has proven to be a reliable and accurate instrument.}
	\label{fig:floatenstein}
\end{figure}

A small-scale, home-produced drifting buoy was built following the v2021 design in the context of the OneOcean expedition, a circumnavigation by the Norwegian tall ship
Statsraad Lehmkuh. The electronics are identical to everything that was discussed above, except for the battery solution used.
Indeed, since the instrument had to be transported by plane in the checked luggage of an expedition member, 2 traditional alkaline D-cell batteries were used
instead of the Li-batteries of the other instruments. A standard rectangular rigid plastic enclosure of size 12cm x 12cm x 9cm was used as the main body of the buoy. Since
alkaline batteries are heavier than lithium batteries, this resulted in too low floatability. As a consequence, we added small chunks of styrofoam (wrapped in duct tape to
protect from abrasion from the elements), on the sides of the rigid plastic enclosure, in order to increase buoyancy. These additional buoyancy elements were set in place
using cable ties and glued in position using bathroom silicone sealant. In addition, the rigid plastic box was fully sealed with epoxy glue, to make it fully watertight.
Finally, a keel composed of diverse scrap metallic parts (spare screws and bolts), packed in duct tape, was fixed at the bottom of the plastic enclosure to improve its
static stability. This combination of disparate building materials gave a very peculiar appearance to the drifting buoy, which got nicknamed humouristically as
"Floatenstein" (Floating-Frankenstein). A picture of the instrument, fully built before being shipped, is presented in Fig. \ref{fig:floatenstein}.

Despite its unusual appearance, Floatenstein proved to be a robust and well-functioning drifter. Floatenstein was released in the open ocean on November 11th 2021, and, at
the time of writing this manuscript (January 2022), Floatenstein has been active for a bit over two months and is working nominally. Data transmitted include i) GPS position
sampled each 30 minutes, and ii) vertical wave spectrum sampled each 2 hours, similar to what is described in Section 2.

In order to validate the good functioning of Floatenstein and the electronics inside, we compare the wave data transmitted by Floatenstein with direct satellite altimeter
measurements. The Wavy software package (see \url{https://github.com/bohlinger/wavy}, \citet{BOHLINGER2019101404}) is used to perform i) gathering of relevant satellite
data, ii) collocation with the position of Floatenstein, iii) extraction of relevant fields. For the collocation process, the temporal and spatial constraints are 30 minutes and 50~km, respectively. Four satellite missions had coinciding satellite measurements Sentinel-3A/B, Jason-3, and the altimeter mounted on CFOSAT. The satellite observations are of level 3 processing quality and publicly available in the Copernicus CMEMS archive\footnote{\url{https://resources.marine.copernicus.eu/product-detail/WAVE_GLO_WAV_L3_SWH_NRT_OBSERVATIONS_014_001/INFORMATION}}.

\begin{figure}[htp]
	\centering
	\includegraphics[width=.65\textwidth]{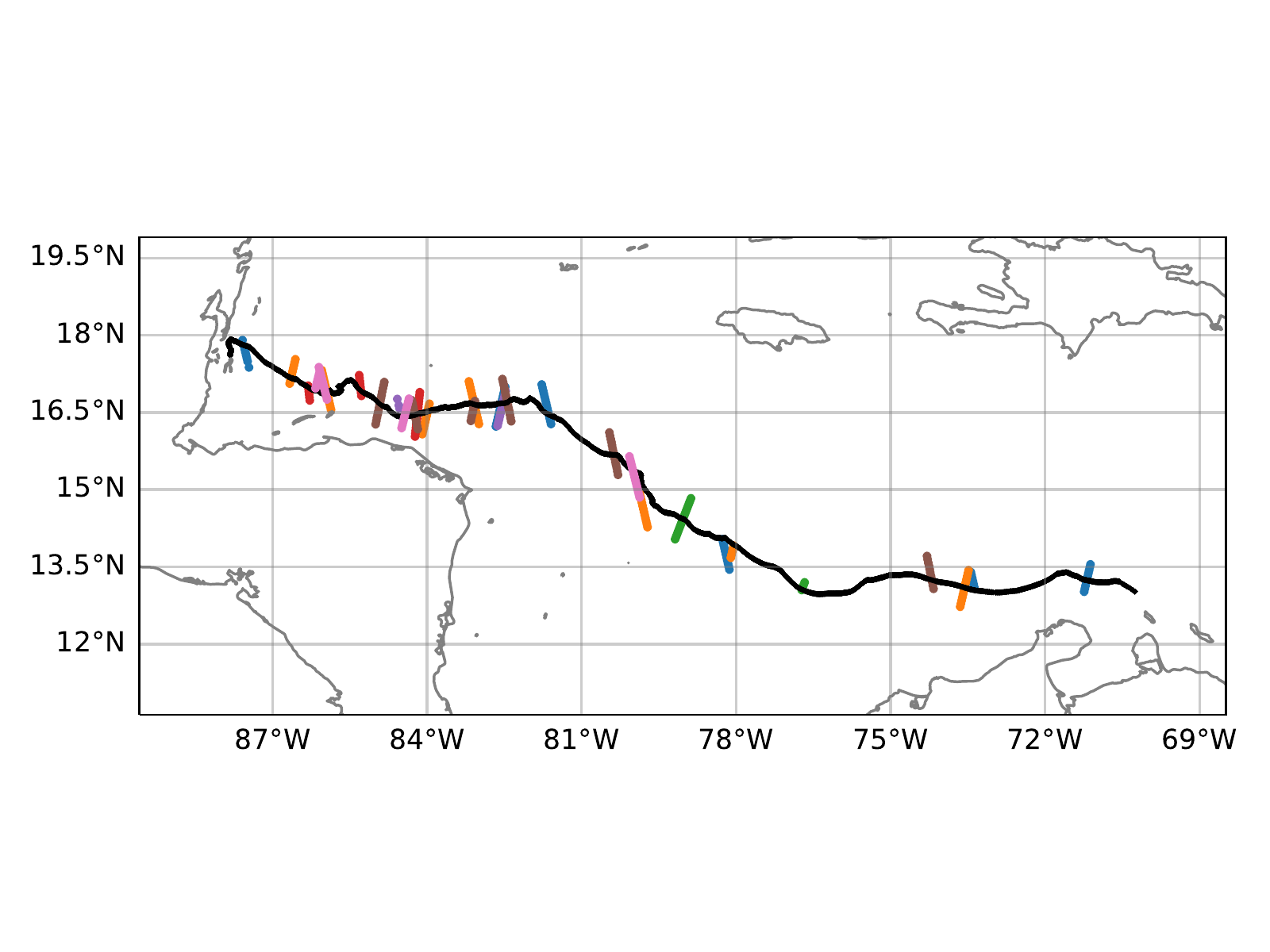}
	\includegraphics[width=.75\textwidth]{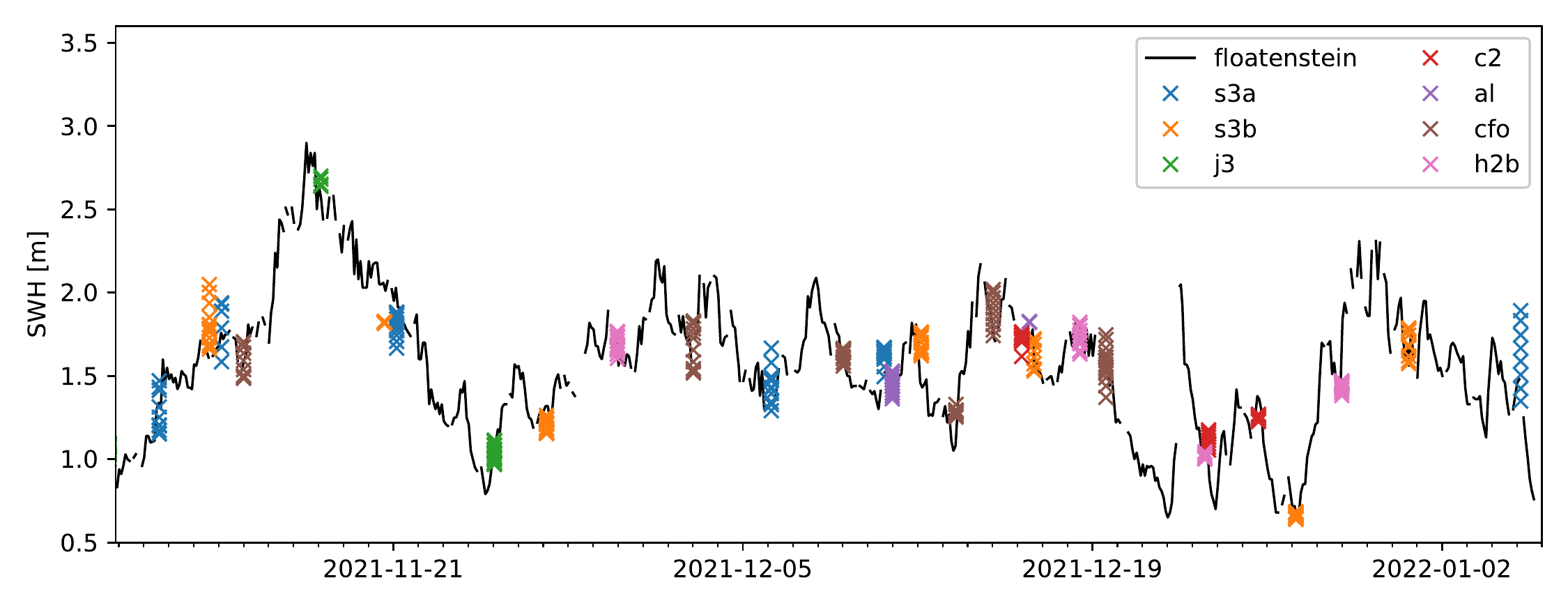}
	\caption{Top: drift pattern of Floatenstein (colored continuous black line), and satellite measurement transects (colored points) over which wave characteristics are measured by the satellites.
Bottom: comparison of the Significant Wave Height (SWH), i) reported by Floatenstein (continuous black line), and ii) from satellite measurements on the corresponding measurement
transects. Excellent agreement is obtained between the data reported by Floatenstein and the satellites.}
	\label{fig:floatenstein_comparison}
\end{figure}

Results of the comparison between Floatenstein and satellite data are presented in Fig. \ref{fig:floatenstein_comparison}. The satellite data are naturally spaced in time,
due to the orbital pattern of the satellites collecting the data. Excellent agreement is observed between the wave statistics reported by Floatenstein, and the direct
observation by the satellites. This is an additional confirmation of the good functioning of the electronics and firmware of the instrument v2021, as well as an illustration
that even coarse drifters are able to accurately measure wave properties, as long as these are small enough to follow the motion of the waves. Indeed, the amplitude of the
hydrodynamic response of the buoy itself is typically a fraction of the size of the buoy. This could, in theory, reach up to several tens of cm for large floating buoys with
a typical size of several meters as is commonly found for commercially available moored system, which has, historically, created a focus on developing wave buoys with a
small hydrodynamic response. Our small buoy, by contrast, has a typical size of 10 cm, which implies that even a "bad" hydrodynamic response, leading to a displacement
response in waves of several tens of percents of the buoys size, is still only a few centimeters in amplitude, well below typical stochastic noise and uncertainty of
stochastic wave measurements in the open ocean.

\section{Conclusion and future work}

This work introduces a new design of waves in ice instrument. This instrument is able to
perform measurements of sea ice drift using a GNSS module, and to measure waves in ice up
to a high degree of accuracy (the noise threshold corresponds to a significant wave height
of around 0.1 cm for waves with period 16 s, as demonstrated in the laboratory experiments), using an inexpensive 9-degree-of-freedom sensor and extensive
signal processing and noise reduction techniques. In addition, this instrument costs only
around slightly under 650 USD in electronics components (including taxes), raw materials, and assembly time, which is around 10 times less
expensive to build than the closest, least expensive commercial alternative we know about (the
Sofar Spotter buoy), while being both more sensitive and much better adapted to working in
the polar night, when solar input is not available. Battery autonomy using a couple of
Lithium D-cells is around 4.5 months, and this can be easily increased by adding more cells
if needed. Assembly time has been cut down to no more than typically half an hour per
instrument, when building a small series of these, thanks to the use of highly integrated open source development boards, and
requires no advanced electronics skills. The code is provided as open source together with
the full design blueprints and assembly instructions. Binary pre-compiled firmware
versions are provided for rapid upload to the electronics boards. Additional sensors can
easily be added to the instruments. Two-way communications are implemented through highly
optimized binary protocols and take place over Iridium, which enables global coverage. All
data measurements take place at fixed time thanks to the real time clock available on
board, to allow easy time synchronization, and data are buffered on board the instrument
to avoid data loss in case of temporary failure in establishing an iridium communication.

This set of characteristics is, as a whole, and to the best of our knowledge, unrivaled by
any commercial or open source instrument. These drastic improvements in cost and power
efficiency, compared even to previous open source instruments such as the one described in \citet{rabault2020open}, are made
possible by the use of a cutting edge smartwatch processor, which can handle all of the
logging and data processing workloads. This is, to the best of our knowledge, the fist
time that such an instrument is built without the need for an expensive, data-hungry
embedded microcomputer, and the first time the full logging, data processing, and
communication workload is controlled from a single, power-efficient, low cost microcontroller. In addition, this design is both modular and adapted to not only measurements
of sea ice drift and waves in ice, but also for open ocean measurements, as we have shown in the validation
section.

We validated the design and signal processing code through a number of test campaigns both
in the laboratory, in the open water, and on sea ice. Results indicate excellent agreement
with state of the art, reference instruments and measurement methods, and confirm the
robustness of the design as well as its power efficiency. In particular, the low power
consumption of the instrument (which is always challenging to get right, as any error in
the code or hardware design can easily increase the sleep power by a few milliamps, which,
while it sounds little, drains battery over time), is confirmed during a deployment in the
Arctic, and the accuracy of the wave measurements and wave data processing code is confirmed
in several test cases.

This work, while purely technical, is in our opinion a groundbreaking step towards
bringing open source, inexpensive, easy to build, modular, power efficient instrumentation
to the field of in-situ study of waves in ice in particular, and oceanographic
measurements in general. Indeed, our design has also proven to be highly efficient both as a
traditional open ocean drifter and wave buoy. We expect that this design will
significantly reduce the barrier to entry for new research groups to monitor waves in ice,
sea ice drift patterns, and similarly wave activity and ocean upper layer drift and
currents in the open ocean, and be a possible platform for public outreach since the
design is simple enough to be assembled for example by high school students. This design
can also be used as a basic building block for setting up a variety of instruments, such
as weather stations, environmental loggers, and wildlife monitoring instruments, by
connecting specific sensors and adding extra modules to the firmware, while taking
advantage of the power efficiency and iridium connectivity of the main electronics board.

Another key improvement brought by the present design is its very limited weight and size.
Typically, our implementations of the design occupy a space of 12cm x 12cm x 9cm and
weight between 0.3 and 0.5kg. This is small and light enough to be deployed for example
from a medium size quadcopter drone, or from a fishing rod several meters long. This could
prove very important in practice when releasing large numbers of drifters in the Arctic
and Antarctic, since i) changing the course of an icebreaker, and forcing it to make many
(even short) stops to put instrumentation on the ice, has a large footprint on the
operation of the ship and a high cost, while ii) flying a quadcopter drone to deploy
instrumentation on the ice, or dropping instrumentation on-the-go while the boat is
steaming using a long rod, would have a zero marginal cost for an expedition.

As a final technical word, we want to say that, while we release the instrument design,
code, and validation data already now since we believe that it may already be of value for
other groups, there are a number of ongoing improvements that are taking place and will
continue also after this paper is published. First, we are working on adding the
possibility to store large amounts of data using a 2-step memory setup (i.e., a
Ferromagnetic RAM for fast, low power storing of data on the fly, and a SD card for slow
long-term archiving). This will make the instrument even more valuable in contexts where
it can be recovered after deployment, since this will allow to store and recover time
series that are, at present, only stored temporarily in RAM for performing wave spectrum
computation. Second, we will work in the future on adding full directional wave spectrum
information. Finally, when these additions are done, a major code rewrite, taking into
account all the lessons learnt while developing this instrument, will be performed to make
the code even easier to extend, test, and reuse. When these further refinements are implemented,
we will also consider designing an open source printed circuit board, to lower even further
the amount of work needed to assemble the electronics of the instrument. Ultimately, if there
is interest from the community, we hope to provide an all-in-one, fully integrated, low cost
electronics boards that will be a true turn-key solution.

We hope that the release of our design may spark the emergence of an ecosystem focused
around open source for polar science and geophysics. In this spirit, we release all the
materials on Github as previously mentioned (see Appendix A for more details), and we
invite any interested reader to engage on our repository there, both asking questions if
anything needs further clarification, discussing further code and hardware improvement, or
simply asking for help getting started with our design if they need some. We believe that
Github is a very exciting platform for such exchanges, thanks to the tight integration
between the issues tracker (which can be used as a tool for chatting and discussion), and
the code and hardware designs. We hope that this can become, over time, a platform for
sharing information, experience, and practical tips, within our community.

Finally, we want to emphasize that this design is the continuation of a growing series of
open source instruments released by authors of the present work, which started from simple
low-cost loggers \citep{rabault2016measurements}, evolved into "traditional" designs including a microcontroller
for logging data and a microcomputer for processing it \citep{rabault2020open}, and now reaching the point
where a bleeding edge microcontroller is able to handle the whole logging and data
analysis workload, with the many advantages we discussed here resulting as a consequence.
We believe that the solution we present now is a much better electronics design than
previous hybrid microcomputer - microcontroller solutions, and will mark a transition in
our community. However, the present design is probably not the "endgame" of
instrumentation development yet. In particular, managing to execute the whole workload on
the (still relatively) limited resources of the current microcontroller requires a bit of technical
proficiency at the moment (though this sensitive part is implemented fully by our code,
and adding extra sensors will be a much easier task for the user). However, due to the
current developments in high performance microcontrollers following the explosion of the
smartwatches and similar wearable devices, we believe that much improved microcontrollers
will be available within 3 to 5 years, at which point a new iteration in this open source
instrument will allow to make the firmware both simpler and even more modular, and will
bring this series of instruments to reach a true "endgame" design from which there will be
little further gains to be envisioned.

As a closing word, we want to stress that our technical solution is enabled primarily thanks to the
emergence of strong open source communities within the domain of electronics, microcontrollers, and high level libraries for interacting with individual electronics
components. A lot of this evolution can be made thanks to the growth of companies making a
busyness out of the development and sale of fairly priced open source frameworks and
boards (we want to mention, in particular, Sparkfun Electronics, Adafruit, Arduino,
and Pololu). In our opinion, this transition from a situation where experts in electronics
and low level programming were needed to effectively use various electronics components
and microcontrollers, towards a situation with widely available open source electronics,
solid ecosystems, and easy-to-use libraries (with the Arduino ecosystem and its
derivatives leading the way), has made it possible for virtually anybody to produce
advanced low cost, power efficient embedded designs.

\section*{Acknowledgements}

We want to thank Pr. Frank Nilsen and Pr. Ilker Fer for inviting us to join the Nansen
Legacy Cruise PC-2: Winter Process Cruise, during which we tested the first early
prototypes of the v2021. Our warmest thanks go to all the participants of this cruise and
to the crew of the R/V Kronprins Haakon for their help during this cruise and the very
friendly weeks spent together on board the ship. Our warmest thanks go to Ceslav Czyz, Zoe Koenig, and
Helge Bryhni, for many interesting discussions about the development of custom made
software and hardware for geoscience measurements.

We also want to thank the Norwegian
Meteorological Institute for continuous support of our
efforts towards building an open source instrumentation community, and for support
of our open source release policy. This work was supported partly through several projects, in
particular, ThinTEC (project number A321200), ThinIce (project number 66017), funded by the Research
Council of Norway and the Norwegian Polar Institute. Participation to the Nansen Legacy PC-2 cruise and instrumentation
used there, as well as part of the data analyzis, were funded through Nansen Legacy project (NFR-276730) and FOCUS project (NFR- 301450). In addition, some synergies with the
Machine Ocean project funded by the research Council of Norway (project number 303411)
helped in the present developments.

We gratefully acknowledge support from the Japan Agency for Marine-Earth Science and Technology, in particular,
for constructing the ice-wave tank at the Univ. of Tokyo. The 2021 NABOS buoy deployment was supported by the ArCS II Japan-Russia-Canada International Exchange Program (https://sites.google.com/edu.k.u-tokyo.ac.jp/arcsii-iep-jrc/main).
TW is grateful to Dr Tatiana Alekseeva of Russian Arctic and Antarctic Research Institute for making our NABOS deployment possible.
TN, TW, TKo, and TKa  are grateful to Profs Kanna and Tateyama for deploying our wave buoys during 2021 NABOS expedition.
Some of this work was performed in the the context of the Arctic Challenge for Sustainability II Project (ArCS II Project, Program Grant Number JPMXD1420318865).
A part of this study was also conducted under JSPS KAKENHI Grant Number JP 19H00801, 19H05512, and 21K14357.
This study was supported partly by the Grant for Joint Research Program of the Japan Arctic Research Network Center.

We also want to thank the One Ocean expedition and the University of Bergen course SDG313 for facilitating deployment of the Floatenstein drifter in the Caribbean.
This work was also partially done in the context of the DOFI Petromaks II project (funding to the University of Oslo, by the RCN funding, grant number 28062).

We want to thank Jim Thomson, as well as the members of his group, for very interesting and stimulating
discussions about open source instrumentation and measurements of ocean waves using open source electronics.

Finally, we want to acknowledge and
thank the Arduino community and ecosystem, as well as Sparkfun and the community around
it, for making high performance microelectronics easily available to non experts. Our
development of open source instrumentation as is presented in this work would not be
possible without the amazing work and tools offered by these communities.

\section*{Appendix A: open source release and founding of an open source community}

All the code, instructions, and post processing scripts necessary to build, program, and
use the instrument v2021, are available as open source hardware and software on Github
under a MIT license at the following URL: \url{https://github.com/jerabaul29/OpenMetBuoy-v2021a}. We
aim at using Github as a central hub for sharing knowledge, tips, asking for new features,
reporting bugs, and driving the development of our instrument. We invite all interested
readers to discuss with us, ask for help, share their experience, and turn this project
into an active community, by engaging with us through the issue tracker available there.

\section*{Appendix B: Zeni-v2021 assembly for the 2021 NABOS cruise}\label{sec:AppB}

\begin{figure}[htp]
	\centering
	\includegraphics[width=.85\textwidth]{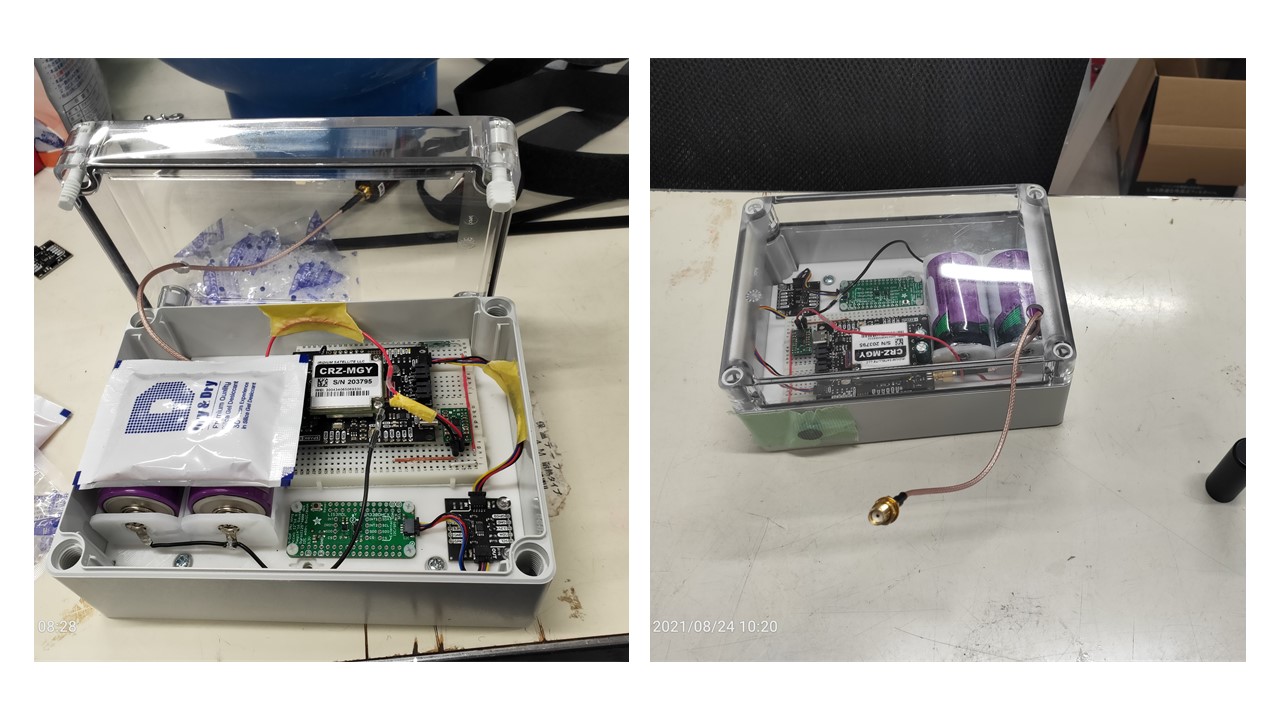}
	\caption{Illustrations of the instrument v2021 electronics components enclosed in a water resistant Takachi box. }
	\label{fig:v2021TakachiBox}
\end{figure}

\begin{figure}[htp]
	\centering
	\includegraphics[width=.85\textwidth]{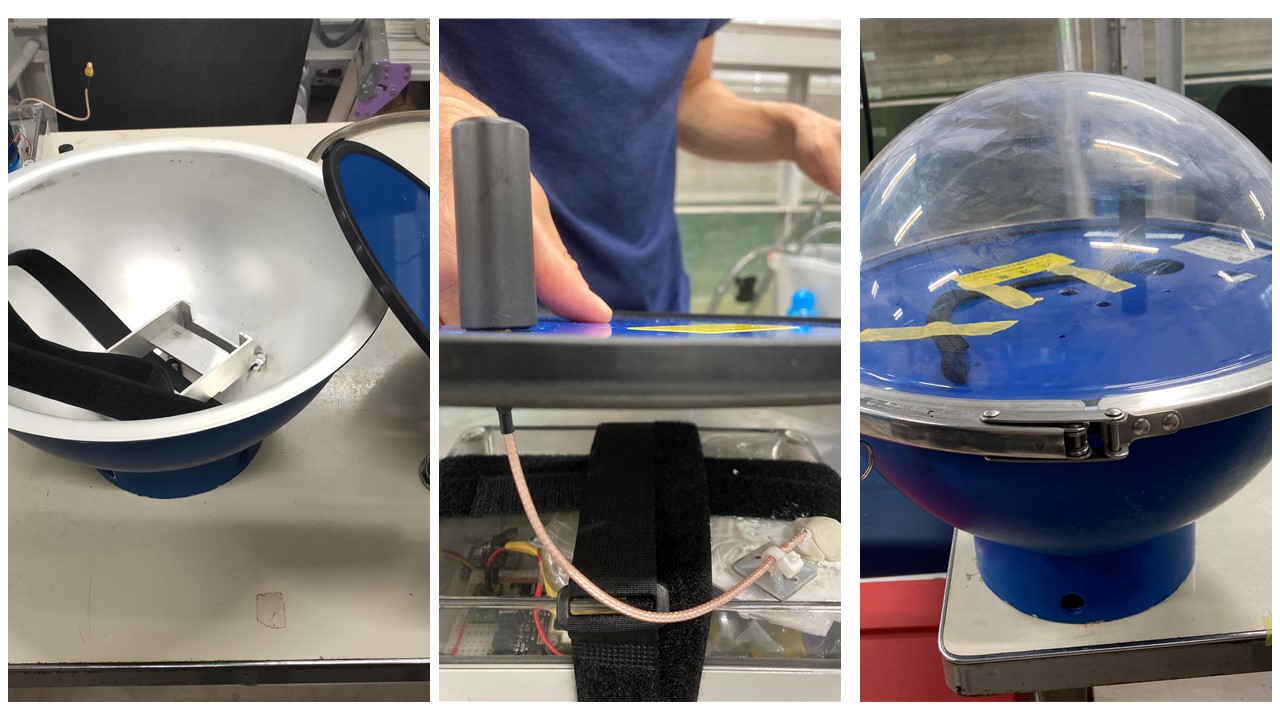}
	\caption{Illustrations of Zeni Lite drifting buoy being used as a floating housing. The left image shows the bottom part of the buoy where the Takachi box was fixed. The middle image shows how the antenna was attached through the blue plate. The right image shows the assembled Zeni-v2021.}
	\label{fig:v2021ZeniHull}
\end{figure}

The
Artemis Global Tracker and other electronics components listed in
Section \ref{subsec:DesignOverview} were enclosed in a water resistant Takachi box as shown
in Fig \ref{fig:v2021TakachiBox}. Deployment in the open water requires a floating
enclosure that is durable and watertight. For our purpose, we used a
commercial Zeni Lite drifting buoy (\url{https://www.zenilite.co.jp/prod/new-chikuden.html}), that was purchased by the University of Tokyo several years ago,
as the floating enclosure.
Images of the floating enclosure setup are shown in Fig \ref{fig:v2021ZeniHull}. The Zeni
Lite drifting buoy has open space inside where the Takachi box was fixed. This space was
easily accessible, and it is also high enough for the antenna to be fixed on the upper half of
the buoy. The water tightness is achieved via an O-ring placed underneath the blue plate.
When the steel brace is fastened, pressure is evenly applied around the buoy housing where
the top and bottom parts of the buoy are joined. The instrument v2021 that was deployed in the
Arctic Ocean as described in Section \ref{subsec:NABOSdeployment}, is referred to as
Zeni-v2021 in the main text, because the instrument v2021 was assembled in the Zeni Lite
drifting buoy enclosure.

\bibliographystyle{ametsoc2014}
\bibliography{references}

\begin{thebibliography}{70}
\providecommand{\natexlab}[1]{#1}
\providecommand{\url}[1]{\texttt{#1}}
\renewcommand{\UrlFont}{\rmfamily}
\providecommand{\urlprefix}{URL }
\expandafter\ifx\csname urlstyle\endcsname\relax
  \providecommand{\doi}[1]{doi:\discretionary{}{}{}#1}\else
  \providecommand{\doi}{doi:\discretionary{}{}{}\begingroup
  \urlstyle{rm}\Url}\fi
\providecommand{\eprint}[2][]{\url{#2}}

\bibitem[{{Ambiq Inc.}(2021)}]{ambiq}
{Ambiq Inc.}, 2021: Ambiq apollo3 blu microcontroller.
  \urlprefix\url{https://ambiq.com/apollo3-blue/}, Web page.

\bibitem[{Ardhuin et~al.(2020)Ardhuin, Otero, Merrifield, Grouazel,, and
  Terrill}]{ardhuin2020ice}
Ardhuin, F., M.~Otero, S.~Merrifield, A.~Grouazel, and E.~Terrill, 2020: Ice
  breakup controls dissipation of wind waves across southern ocean sea ice.
  \textit{Geophysical Research Letters}, \textbf{47~(13)}, e2020GL087\,699.

\bibitem[{Ardhuin et~al.(2017)}]{ARDHUIN2017211}
Ardhuin, F., and Coauthors, 2017: Measuring ocean waves in sea ice using sar
  imagery: A quasi-deterministic approach evaluated with sentinel-1 and in situ
  data. \textit{Remote Sensing of Environment}, \textbf{189}, 211--222,
  \doi{https://doi.org/10.1016/j.rse.2016.11.024},
  \urlprefix\url{https://www.sciencedirect.com/science/article/pii/S0034425716304710}.

\bibitem[{Batrak and M{\"u}ller(2018)Batrak, and
  M{\"u}ller}]{batrak2018atmospheric}
Batrak, Y., and M.~M{\"u}ller, 2018: Atmospheric response to kilometer-scale
  changes in sea ice concentration within the marginal ice zone.
  \textit{Geophysical Research Letters}, \textbf{45~(13)}, 6702--6709.

\bibitem[{Bohlinger et~al.(2019)Bohlinger, Øyvind Breivik, Economou,, and
  Müller}]{BOHLINGER2019101404}
Bohlinger, P., Øyvind Breivik, T.~Economou, and M.~Müller, 2019: A novel
  approach to computing super observations for probabilistic wave model
  validation. \textit{Ocean Modelling}, \textbf{139}, 101\,404,
  \doi{https://doi.org/10.1016/j.ocemod.2019.101404},
  \urlprefix\url{https://www.sciencedirect.com/science/article/pii/S1463500319300435}.

\bibitem[{Cheng et~al.(2017)}]{https://doi.org/10.1002/2017JC013275}
Cheng, S., and Coauthors, 2017: Calibrating a viscoelastic sea ice model for
  wave propagation in the arctic fall marginal ice zone. \textit{Journal of
  Geophysical Research: Oceans}, \textbf{122~(11)}, 8770--8793,
  \doi{https://doi.org/10.1002/2017JC013275},
  \urlprefix\url{https://agupubs.onlinelibrary.wiley.com/doi/abs/10.1002/2017JC013275},
  \eprint{https://agupubs.onlinelibrary.wiley.com/doi/pdf/10.1002/2017JC013275}.

\bibitem[{{Datawell Corporation}(2001)}]{DataWellHistory}
{Datawell Corporation}, 2001: {H}istory of {D}atawell.
  \url{https://www.datawell.nl/Portals/0/Documents/Brochures/datawell_brochure_history_b-13-01.pdf},
  accessed 2/11/21.

\bibitem[{{Defense Advanced Research Projects Agency}(2021)}]{darpa_oot}
{Defense Advanced Research Projects Agency}, 2021: Ocean of things.
  \urlprefix\url{https://oceanofthings.darpa.mil/}, Web page.

\bibitem[{Doble et~al.(2006)Doble, Mercer, Meldrum,, and Peppe}]{doble2006wave}
Doble, M., D.~J. Mercer, D.~Meldrum, and O.~C. Peppe, 2006: Wave measurements
  on sea ice: developments in instrumentation. \textit{Annals of Glaciology},
  \textbf{44}, 108--112.

\bibitem[{Golden et~al.(2020)}]{golden2020modeling}
Golden, K.~M., and Coauthors, 2020: Modeling sea ice. \textit{Notices of the
  American Mathematical Society}, \textbf{67~(10)}.

\bibitem[{Herman(2010)}]{herman2010sea}
Herman, A., 2010: Sea-ice floe-size distribution in the context of spontaneous
  scaling emergence in stochastic systems. \textit{Physical Review E},
  \textbf{81~(6)}, 066\,123.

\bibitem[{Herman(2017)}]{herman2017wave}
Herman, A., 2017: Wave-induced stress and breaking of sea ice in a coupled
  hydrodynamic discrete-element wave--ice model. \textit{The Cryosphere},
  \textbf{11~(6)}, 2711--2725.

\bibitem[{Herman(2018)}]{herman2018wave}
Herman, A., 2018: Wave-induced surge motion and collisions of sea ice floes:
  Finite-floe-size effects. \textit{Journal of Geophysical Research: Oceans},
  \textbf{123~(10)}, 7472--7494.

\bibitem[{Herman(2021)}]{herman2021spectral}
Herman, A., 2021: Spectral wave energy dissipation due to under-ice turbulence.
  \textit{Journal of Physical Oceanography}, \textbf{51~(4)}, 1177--1186.

\bibitem[{Herman et~al.(2018)Herman, Evers,, and Reimer}]{herman2018floe}
Herman, A., K.-U. Evers, and N.~Reimer, 2018: Floe-size distributions in
  laboratory ice broken by waves. \textit{The Cryosphere}, \textbf{12~(2)},
  685--699.

\bibitem[{Herman et~al.(2021)Herman, Wenta,, and Cheng}]{herman2021sizes}
Herman, A., M.~Wenta, and S.~Cheng, 2021: Sizes and shapes of sea ice floes
  broken by waves--a case study from the east antarctic coast.
  \textit{Frontiers in Earth Science}, \textbf{9}, 390.

\bibitem[{Hori et~al.(2012)Hori, Yabuki, Sugimura,, and Terui}]{ADS-AMSR2}
Hori, M., H.~Yabuki, T.~Sugimura, and T.~Terui, 2012: {AMSR2 Level 3} product
  of daily polar brightness temperatures and product, 1.00. Arctic Data archive
  System (ADS), Japan,
  \urlprefix\url{https://ads.nipr.ac.jp/dataset/A20170123-003}, [Date accessed
  23. Jul. 2019].

\bibitem[{Horvat et~al.(2020)Horvat, Blanchard-Wrigglesworth,, and
  Petty}]{horvat2020observing}
Horvat, C., E.~Blanchard-Wrigglesworth, and A.~Petty, 2020: Observing waves in
  sea ice with icesat-2. \textit{Geophysical Research Letters},
  \textbf{47~(10)}, e2020GL087\,629.

\bibitem[{Horvat and Tziperman(2015)Horvat, and
  Tziperman}]{horvat2015prognostic}
Horvat, C., and E.~Tziperman, 2015: A prognostic model of the sea-ice floe size
  and thickness distribution. \textit{The Cryosphere}, \textbf{9~(6)},
  2119--2134.

\bibitem[{Horvat and Tziperman(2017)Horvat, and
  Tziperman}]{horvat2017evolution}
Horvat, C., and E.~Tziperman, 2017: The evolution of scaling laws in the sea
  ice floe size distribution. \textit{Journal of Geophysical Research: Oceans},
  \textbf{122~(9)}, 7630--7650.

\bibitem[{Johnson et~al.(2021)Johnson, Marchenko, Dammann,, and
  Mahoney}]{johnson2021observing}
Johnson, M.~A., A.~V. Marchenko, D.~O. Dammann, and A.~R. Mahoney, 2021:
  Observing wind-forced flexural-gravity waves in the beaufort sea and their
  relationship to sea ice mechanics. \textit{Journal of Marine Science and
  Engineering}, \textbf{9~(5)}, 471.

\bibitem[{Keating(2016)}]{keating2016fetch}
Keating, D., 2016: Fetch-limited wave growth in nootka sound.

\bibitem[{Kodaira et~al.(2021)Kodaira, Waseda, Nose, Sato, Inoue, Voermans,,
  and Babanin}]{KODAIRA2021100567}
Kodaira, T., T.~Waseda, T.~Nose, K.~Sato, J.~Inoue, J.~Voermans, and
  A.~Babanin, 2021: Observation of on-ice wind waves under grease ice in the
  western arctic ocean. \textit{Polar Science}, \textbf{27}, 100\,567,
  \doi{https://doi.org/10.1016/j.polar.2020.100567},
  \urlprefix\url{https://www.sciencedirect.com/science/article/pii/S1873965220300761},
  arctic Challenge for Sustainability Project (ArCS).

\bibitem[{Kohout et~al.(2015)Kohout, Penrose, Penrose,, and
  Williams}]{kohout2015device}
Kohout, A.~L., B.~Penrose, S.~Penrose, and M.~J. Williams, 2015: A device for
  measuring wave-induced motion of ice floes in the antarctic marginal ice
  zone. \textit{Annals of Glaciology}, \textbf{56~(69)}, 415--424.

\bibitem[{Kohout et~al.(2020)Kohout, Smith, Roach, Williams, Montiel,, and
  Williams}]{kohout_smith_roach_williams_montiel_williams_2020}
Kohout, A.~L., M.~Smith, L.~A. Roach, G.~Williams, F.~Montiel, and M.~J.~M.
  Williams, 2020: Observations of exponential wave attenuation in antarctic sea
  ice during the pipers campaign. \textit{Annals of Glaciology},
  \textbf{61~(82)}, 196–209, \doi{10.1017/aog.2020.36}.

\bibitem[{Li et~al.(2021{\natexlab{a}})Li, Gedikli,, and
  Lubbad}]{Li2021LaboratorySO}
Li, H., E.~D. Gedikli, and R.~Lubbad, 2021{\natexlab{a}}: Laboratory study of
  wave-induced flexural motion of ice floes. \textit{Cold Regions Science and
  Technology}, \textbf{182}, 103\,208.

\bibitem[{Li et~al.(2021{\natexlab{b}})Li, Babanin, Liu, Voermans, Heil,, and
  Tang}]{li2021effects}
Li, J., A.~V. Babanin, Q.~Liu, J.~J. Voermans, P.~Heil, and Y.~Tang,
  2021{\natexlab{b}}: Effects of wave-induced sea ice break-up and mixing in a
  high-resolution coupled ice-ocean model. \textit{Journal of Marine Science
  and Engineering}, \textbf{9~(4)}, 365.

\bibitem[{L{\o}ken et~al.(2021{\natexlab{a}})L{\o}ken, Ellevold, de~la Torre,
  Rabault,, and Jensen}]{loken2021bringing}
L{\o}ken, T.~K., T.~J. Ellevold, R.~G.~R. de~la Torre, J.~Rabault, and
  A.~Jensen, 2021{\natexlab{a}}: Bringing optical fluid motion analysis to the
  field: a methodology using an open source rov as a camera system and rising
  bubbles as tracers. \textit{Measurement Science and Technology},
  \textbf{32~(9)}, 095\,302.

\bibitem[{L{\o}ken et~al.(2021{\natexlab{b}})L{\o}ken, Marchenko, Ellevold,
  Rabault,, and Jensen}]{loken2021investigation}
L{\o}ken, T.~K., A.~Marchenko, T.~J. Ellevold, J.~Rabault, and A.~Jensen,
  2021{\natexlab{b}}: An investigation into the turbulence induced by moving
  ice floes. \textit{arXiv preprint arXiv:2104.02378}.

\bibitem[{L{\o}ken et~al.(2021{\natexlab{c}})L{\o}ken, Rabault, Jensen,
  Sutherland, Christensen,, and M{\"u}ller}]{loken2021wave}
L{\o}ken, T.~K., J.~Rabault, A.~Jensen, G.~Sutherland, K.~H. Christensen, and
  M.~M{\"u}ller, 2021{\natexlab{c}}: Wave measurements from ship mounted
  sensors in the arctic marginal ice zone. \textit{Cold Regions Science and
  Technology}, \textbf{182}, 103\,207.

\bibitem[{Marchenko et~al.(2021)Marchenko, Haase, Jensen, Lishman, Rabault,
  Evers, Shortt,, and Thiel}]{marchenko2021laboratory}
Marchenko, A., A.~Haase, A.~Jensen, B.~Lishman, J.~Rabault, K.-U. Evers,
  M.~Shortt, and T.~Thiel, 2021: Laboratory investigations of the bending
  rheology of floating saline ice and physical mechanisms of wave damping in
  the hsva hamburg ship model basin ice tank. \textit{Water}, \textbf{13~(8)},
  1080.

\bibitem[{Marchenko et~al.(2017)Marchenko, Rabault, Sutherland, Collins,
  Wadhams,, and Chumakov}]{marchenko2017field}
Marchenko, A., J.~Rabault, G.~Sutherland, C.~O. Collins, P.~Wadhams, and
  M.~Chumakov, 2017: Field observations and preliminary investigations of a
  wave event in solid drift ice in the barents sea.
  \textit{Proceedings-International Conference on Port and Ocean Engineering
  under Arctic Conditions}, Port and Ocean Engineering under Arctic Conditions.

\bibitem[{Marchenko et~al.(2019)Marchenko, Wadhams, Collins, Rabault,, and
  Chumakov}]{marchenko2019wave}
Marchenko, A., P.~Wadhams, C.~Collins, J.~Rabault, and M.~Chumakov, 2019:
  Wave-ice interaction in the north-west barents sea. \textit{Applied Ocean
  Research}, \textbf{90}, 101\,861.

\bibitem[{Mosig et~al.(2015)Mosig, Montiel,, and
  Squire}]{https://doi.org/10.1002/2015JC010881}
Mosig, J. E.~M., F.~Montiel, and V.~A. Squire, 2015: Comparison of
  viscoelastic-type models for ocean wave attenuation in ice-covered seas.
  \textit{Journal of Geophysical Research: Oceans}, \textbf{120~(9)},
  6072--6090, \doi{https://doi.org/10.1002/2015JC010881},
  \urlprefix\url{https://agupubs.onlinelibrary.wiley.com/doi/abs/10.1002/2015JC010881},
  \eprint{https://agupubs.onlinelibrary.wiley.com/doi/pdf/10.1002/2015JC010881}.

\bibitem[{{Mouser Inc.}(2021)}]{reed_switch}
{Mouser Inc.}, 2021: Mdrr-dt-15-25-f reed switch.
  \urlprefix\url{https://no.mouser.com/ProductDetail/Littelfuse/MDRR-DT-15-25-F?qs=nyo4TFax6Nfw7Od4323OuQ%3D%3D},
  Web page.

\bibitem[{Nilsen et~al.(2021)}]{nilsen2021pc}
Nilsen, F., and Coauthors, 2021: Pc-2 winter process cruise (wpc): Cruise
  report. \textit{The Nansen Legacy Report Series}, \textbf{~(26)}.

\bibitem[{Nose et~al.(2018)Nose, Webb, Waseda, Inoue,, and Sato}]{Nose2018}
Nose, T., A.~Webb, T.~Waseda, J.~Inoue, and K.~Sato, 2018: Predictability of
  storm wave heights in the ice-free {Beaufort Sea}. \textit{Ocean Dynamics},
  \textbf{68~(10)}, 1383--1402,
  \url{https://doi.org/10.1007/s10236-018-1194-0}.

\bibitem[{Parkinson et~al.(1997)Parkinson, Claire et~al.}]{parkinson1997earth}
Parkinson, C., L.~Claire, and Coauthors, 1997: \textit{Earth from above: using
  color-coded satellite images to examine the global environment}. University
  science books.

\bibitem[{{Pololu Inc.}(2021)}]{pololu}
{Pololu Inc.}, 2021: Pololu 3.3v step-up/step-down voltage regulator s7v8f3.
  \urlprefix\url{https://www.pololu.com/product/2122}, Web page.

\bibitem[{Rabault et~al.(2017)Rabault, Sutherland, Gundersen,, and
  Jensen}]{rabault2017measurements}
Rabault, J., G.~Sutherland, O.~Gundersen, and A.~Jensen, 2017: Measurements of
  wave damping by a grease ice slick in svalbard using off-the-shelf sensors
  and open-source electronics. \textit{Journal of Glaciology},
  \textbf{63~(238)}, 372--381.

\bibitem[{Rabault et~al.(2020)Rabault, Sutherland, Gundersen, Jensen,
  Marchenko,, and Breivik}]{rabault2020open}
Rabault, J., G.~Sutherland, O.~Gundersen, A.~Jensen, A.~Marchenko, and
  {\O}.~Breivik, 2020: An open source, versatile, affordable waves in ice
  instrument for scientific measurements in the polar regions. \textit{Cold
  Regions Science and Technology}, \textbf{170}, 102\,955.

\bibitem[{Rabault et~al.(2019)Rabault, Sutherland, Jensen, Christensen,, and
  Marchenko}]{rabault2019experiments}
Rabault, J., G.~Sutherland, A.~Jensen, K.~H. Christensen, and A.~Marchenko,
  2019: Experiments on wave propagation in grease ice: combined wave gauges and
  particle image velocimetry measurements. \textit{Journal of Fluid Mechanics},
  \textbf{864}, 876--898.

\bibitem[{Rabault et~al.(2016)Rabault, Sutherland, Ward, Christensen, Halsne,,
  and Jensen}]{rabault2016measurements}
Rabault, J., G.~Sutherland, B.~Ward, K.~H. Christensen, T.~Halsne, and
  A.~Jensen, 2016: Measurements of waves in landfast ice using inertial motion
  units. \textit{IEEE Transactions on Geoscience and Remote Sensing},
  \textbf{54~(11)}, 6399--6408.

\bibitem[{Raghukumar et~al.(2019)Raghukumar, Chang, Spada,, and
  Jannsen}]{raghukumar2019directional}
Raghukumar, K., G.~Chang, F.~Spada, and T.~Jannsen, 2019: Directional spectrum
  measurements by the spotter: a new developed wave buoy.

\bibitem[{Roach et~al.(2019)Roach, Bitz, Horvat,, and Dean}]{roach2019advances}
Roach, L.~A., C.~M. Bitz, C.~Horvat, and S.~M. Dean, 2019: Advances in modeling
  interactions between sea ice and ocean surface waves. \textit{Journal of
  Advances in Modeling Earth Systems}, \textbf{11~(12)}, 4167--4181.

\bibitem[{Roach et~al.(2018)Roach, Horvat, Dean,, and Bitz}]{roach2018emergent}
Roach, L.~A., C.~Horvat, S.~M. Dean, and C.~M. Bitz, 2018: An emergent sea ice
  floe size distribution in a global coupled ocean-sea ice model.
  \textit{Journal of Geophysical Research: Oceans}, \textbf{123~(6)},
  4322--4337.

\bibitem[{{Rockblock7 Inc.}(2021)}]{iridium_power}
{Rockblock7 Inc.}, 2021: Power consumption guidance for the iridium 9603 modem.
  \urlprefix\url{https://docs.rockblock.rock7.com/docs/power-consumption-guidance},
  Web page.

\bibitem[{Smit et~al.(2021)Smit, Houghton, Jordanova, Portwood, Shapiro, Clark,
  Sosa,, and Janssen}]{smit2021assimilation}
Smit, P., I.~Houghton, K.~Jordanova, T.~Portwood, E.~Shapiro, D.~Clark,
  M.~Sosa, and T.~Janssen, 2021: Assimilation of significant wave height from
  distributed ocean wave sensors. \textit{Ocean Modelling}, \textbf{159},
  101\,738.

\bibitem[{Smith et~al.(2018)Smith, Korobkin, Parau, Feltham,, and
  Squire}]{smith2018modelling}
Smith, F., A.~Korobkin, E.~Parau, D.~Feltham, and V.~Squire, 2018: Modelling of
  sea-ice phenomena. The Royal Society Publishing.

\bibitem[{Smith and Thomson(2020)Smith, and Thomson}]{smith2020pancake}
Smith, M., and J.~Thomson, 2020: Pancake sea ice kinematics and dynamics using
  shipboard stereo video. \textit{Annals of Glaciology}, \textbf{61~(82)},
  1--11.

\bibitem[{{Sparkfun Inc.}(2021)}]{agt}
{Sparkfun Inc.}, 2021: Artemis global tracker.
  \urlprefix\url{https://www.sparkfun.com/products/16469}, Web page.

\bibitem[{Squire(2020)}]{squire2020ocean}
Squire, V.~A., 2020: Ocean wave interactions with sea ice: a reappraisal.
  \textit{Annual Review of Fluid Mechanics}, \textbf{52}, 37--60.

\bibitem[{Sree et~al.(2020{\natexlab{a}})Sree, Law,, and Shen}]{Sree2020AnES}
Sree, D.~K., A.~W.-K. Law, and H.~H. Shen, 2020{\natexlab{a}}: An experimental
  study of gravity waves through segmented floating viscoelastic covers.
  \textit{Applied Ocean Research}, \textbf{101}, 102\,233.

\bibitem[{Sree et~al.(2020{\natexlab{b}})Sree, Law,, and Shen}]{SREE2020102233}
Sree, D.~K., A.~W.-K. Law, and H.~H. Shen, 2020{\natexlab{b}}: An experimental
  study of gravity waves through segmented floating viscoelastic covers.
  \textit{Applied Ocean Research}, \textbf{101}, 102\,233,
  \doi{https://doi.org/10.1016/j.apor.2020.102233},
  \urlprefix\url{https://www.sciencedirect.com/science/article/pii/S0141118719309010}.

\bibitem[{Sutherland et~al.(2021)Sutherland, Aguiar, Hole, Rabault, Dabboor,,
  and Breivik}]{sutherland2021determining}
Sutherland, G., V.~Aguiar, L.-R. Hole, J.~Rabault, M.~Dabboor, and
  {\O}.~Breivik, 2021: Determining an optimal transport velocity in the
  marginal ice zone using operational ice-ocean prediction systems.
  \textit{ArXiv}.

\bibitem[{Sutherland et~al.(2017)Sutherland, Halsne, Rabault,, and
  Jensen}]{sutherland2017attenuation}
Sutherland, G., T.~Halsne, J.~Rabault, and A.~Jensen, 2017: The attenuation of
  monochromatic surface waves due to the presence of an inextensible cover.
  \textit{Wave motion}, \textbf{68}, 88--96.

\bibitem[{Sutherland and Rabault(2016)Sutherland, and
  Rabault}]{sutherland2016observations}
Sutherland, G., and J.~Rabault, 2016: Observations of wave dispersion and
  attenuation in landfast ice. \textit{Journal of Geophysical Research:
  Oceans}, \textbf{121~(3)}, 1984--1997.

\bibitem[{Sutherland et~al.(2019)Sutherland, Rabault, Christensen,, and
  Jensen}]{sutherland2019two}
Sutherland, G., J.~Rabault, K.~H. Christensen, and A.~Jensen, 2019: A two layer
  model for wave dissipation in sea ice. \textit{Applied Ocean Research},
  \textbf{88}, 111--118.

\bibitem[{Thomson(2012)}]{thomson2012wave}
Thomson, J., 2012: Wave breaking dissipation observed with “swift”
  drifters. \textit{Journal of Atmospheric and Oceanic Technology},
  \textbf{29~(12)}, 1866--1882.

\bibitem[{Thomson et~al.(2019{\natexlab{a}})Thomson, Gemmrich, Rogers,
  Collins,, and Ardhuin}]{https://doi.org/10.1029/2019JC015354}
Thomson, J., J.~Gemmrich, W.~E. Rogers, C.~O. Collins, and F.~Ardhuin,
  2019{\natexlab{a}}: Wave groups observed in pancake sea ice. \textit{Journal
  of Geophysical Research: Oceans}, \textbf{124~(11)}, 7400--7411,
  \doi{https://doi.org/10.1029/2019JC015354},
  \urlprefix\url{https://agupubs.onlinelibrary.wiley.com/doi/abs/10.1029/2019JC015354},
  \eprint{https://agupubs.onlinelibrary.wiley.com/doi/pdf/10.1029/2019JC015354}.

\bibitem[{Thomson and Rogers(2014)Thomson, and
  Rogers}]{https://doi.org/10.1002/2014GL059983}
Thomson, J., and W.~E. Rogers, 2014: Swell and sea in the emerging arctic
  ocean. \textit{Geophysical Research Letters}, \textbf{41~(9)}, 3136--3140,
  \doi{https://doi.org/10.1002/2014GL059983},
  \urlprefix\url{https://agupubs.onlinelibrary.wiley.com/doi/abs/10.1002/2014GL059983},
  \eprint{https://agupubs.onlinelibrary.wiley.com/doi/pdf/10.1002/2014GL059983}.

\bibitem[{Thomson et~al.(2016)}]{thomson2016emerging}
Thomson, J., and Coauthors, 2016: Emerging trends in the sea state of the
  beaufort and chukchi seas. \textit{Ocean Modelling}, \textbf{105}, 1--12.

\bibitem[{Thomson et~al.(2019{\natexlab{b}})}]{thomson2019new}
Thomson, J., and Coauthors, 2019{\natexlab{b}}: A new version of the swift
  platform for waves, currents, and turbulence in the ocean surface layer.
  \textit{2019 IEEE/OES Twelfth Current, Waves and Turbulence Measurement
  (CWTM)}, IEEE, 1--7.

\bibitem[{Voermans et~al.(2019)Voermans, Babanin, Thomson, Smith,, and
  Shen}]{voermans2019wave}
Voermans, J., A.~Babanin, J.~Thomson, M.~Smith, and H.~Shen, 2019: Wave
  attenuation by sea ice turbulence. \textit{Geophysical Research Letters},
  \textbf{46~(12)}, 6796--6803.

\bibitem[{Voermans et~al.(2020)}]{voermans2020experimental}
Voermans, J.~J., and Coauthors, 2020: Experimental evidence for a universal
  threshold characterizing wave-induced sea ice break-up. \textit{The
  Cryosphere}, \textbf{14~(11)}, 4265--4278.

\bibitem[{Voermans et~al.(2021)}]{tc-2021-210}
Voermans, J.~J., and Coauthors, 2021: Wave dispersion and dissipation in
  landfast ice: comparison of observations against models. \textit{The
  Cryosphere}, \textbf{2021}, 1--26, \doi{10.5194/tc-2021-210},
  \urlprefix\url{https://tc.copernicus.org/preprints/tc-2021-210/}.

\bibitem[{Waseda et~al.(2017)Waseda, Webb, Sato,, and Inoue}]{Waseda2017}
Waseda, T., A.~Webb, K.~Sato, and J.~Inoue, 2017: Arctic wave observation by
  drifting type wave buoys in 2016. \textit{The 27th International Ocean and
  Polar Engineering Conference}, San Francisco, California, USA, International
  Society of Offshore and Polar Engineers.

\bibitem[{Wilkinson et~al.(2007)Wilkinson, Wadke, Meldrum, Mercer, Doble,, and
  Wadhams}]{wilkinson2007autonomous}
Wilkinson, J., P.~Wadke, D.~Meldrum, D.~Mercer, M.~Doble, and P.~Wadhams, 2007:
  The autonomous measurement of waves propagating across the arctic ocean.
  \textit{OCEANS 2007}, IEEE, 1--7.

\bibitem[{Williams et~al.(2017)Williams, Rampal,, and
  Bouillon}]{williams2017wave}
Williams, T.~D., P.~Rampal, and S.~Bouillon, 2017: Wave--ice interactions in
  the nextsim sea-ice model. \textit{The Cryosphere}, \textbf{11~(5)},
  2117--2135.

\bibitem[{Zhao and Shen(2018)Zhao, and Shen}]{zhao2018three}
Zhao, X., and H.~H. Shen, 2018: Three-layer viscoelastic model with eddy
  viscosity effect for flexural-gravity wave propagation through ice cover.
  \textit{Ocean Modelling}, \textbf{131}, 15--23.

\end{thebibliography}

\end{document}